\documentclass[aps,prb,twocolumn,footinbib,superscriptaddress,noindent]{revtex4-1}
\usepackage{lipsum}
\usepackage{microtype}
\usepackage{inputenc}
\usepackage{hyperref}
\usepackage{amsmath}
\usepackage{mathtools}
\usepackage{amsfonts}
\usepackage{amssymb}
\usepackage[T1]{fontenc}
\usepackage{hhline}
\usepackage{lmodern}
\usepackage{braket}
\usepackage{array}
\usepackage{hyperref}
\hypersetup{
    colorlinks=true,
    citecolor = blue,
    citebordercolor = {0 1 0}
}
\newcommand{\+}{\dagger }

\begin{document}
\begin{abstract}

We obtain a complete and exact in the weak-coupling limit ($U \rightarrow 0$) ground state phase diagram of the repulsive fermionic Hubbard model on the square lattice for filling factors $0 < n < 2$ and next-nearest-neighbour hopping amplitudes $0 \le t^{\prime} \le 0.5$. Phases are distinguished by the symmetry and the number of nodes of the superfluid order parameter. The phase diagram is richer than may be expected and typically features states with a high --- higher than that of the fundamental mode of the corresponding irreducible representation --- number of nodes. The effective coupling strength in the Cooper channel $\lambda$, which determines the critical temperature $T_c$ of the superfluid transition, is calculated in the whole parameter space and regions with high values of $\lambda$ are identified. It is shown that besides the expected increase of $\lambda$ near the Van Hove singularity line, joining the ferromagnetic and antiferromagnetic points, another region with high values of $\lambda$ can be found at quarter filling and $t^{\prime}=0.5$ due to the presence of a line of nesting at $t^{\prime} \ge 0.5$. The results can serve as benchmarks for controlled non-perturbative methods and guide the ongoing search for high-$T_c$ superconductivity in the Hubbard model.   
 
\end{abstract}

\title{Ground state phase diagram of the repulsive fermionic $t-t^{\prime}$ Hubbard model on the square lattice from weak-coupling}

\author{Fedor \v{S}imkovic}
\affiliation{Department of Physics, King's College London, Strand, London WC2R 2LS, UK}
\author{Xuan-Wen Liu}
\affiliation{Hefei National Laboratory for Physical Sciences at Microscale, Department of Modern Physics,
and Synergetic Innovation Center of Quantum Information and Quantum Physics,
University of Science and Technology of China, Hefei, Anhui 230026, China}
\author{Youjin Deng}
\affiliation{Hefei National Laboratory for Physical Sciences at Microscale, Department of Modern Physics,
and Synergetic Innovation Center of Quantum Information and Quantum Physics,
University of Science and Technology of China, Hefei, Anhui 230026, China}
\author{Evgeny Kozik}
\affiliation{Department of Physics, King's College London, Strand, London WC2R 2LS, UK}

\bibliographystyle{unsrtnat}

\maketitle


\section{Introduction}

 The repulsive Hubbard model \cite{hubbard1963electron, anderson1987resonating}
\begin{align}
\operatorname{H} &= -t \! \! \sum_{\left< i,j\right>,\sigma} \! \! \hat{c}^{\+}_{i\sigma} \hat{c}^{}_{j\sigma} + t^{\prime}\! \! \! \! \sum_{\left<\left< i,j\right>\right>,\sigma} \! \! \! \! \hat{c}^{\+}_{i\sigma} \hat{c}^{}_{j\sigma} + U \sum_{i} \hat{n}^{}_{i\uparrow} \hat{n}^{}_{i \downarrow} -\mu \sum_{i,\sigma} \hat{n}_{i\sigma},
\label{hubbard}
\end{align}
where $\hat{c}^{\+}_{i\sigma}$ creates a fermion with spin $\sigma = \left\{ \uparrow, \downarrow\right\}$ on the lattice site $i$, $\hat{n}^{}_{i\sigma} = \hat{c}^{\+}_{i\sigma} \hat{c}^{}_{i\sigma}$, $\left< \hdots \right>$  and $\left< \left< \hdots \right> \right>$ denote summation over nearest and next nearest neighbours respectively, $t$ and $t^{\prime}$ are the hopping amplitudes, $U$ the on-site repulsion, and $\mu$ the chemical potential, is widely regarded as paradigmatic for strongly correlated electrons \cite{anderson1997theory, miyake1986, beal1986,scalapino1986, scalapino1999}. It is expected to capture a variety of intriguing macroscopic quantum phenomena, including, e.g., Mott-insulator physics, antiferromagnetism, striped phases, itinerant ferromagnetism, and high-temperature superconductivity. Due to recent remarkable progress in experimental technique, the Hubbard model can be now reliably emulated by ultracold atoms in optical lattices \cite{schneider2008metallic, kohl2005fermionic, jordens2008mott, schneider2012, bakr2009, hofstetter2002, jordens2010} and probed with unprecedented control, which in principle allows to determine its phase diagram experimentally.  

On the theoretical side, the model can be solved exactly in one dimension \cite{sriram1986exact}. Already in $2D$, more relevant in the context of condensed matter systems, obtaining the phase diagram for generic filling factors $n$ and values of the interaction $U$ remains a prohibitively complex problem. Since the seminal work by Kohn and Luttinger\cite{kohn1965}, who showed that the Cooper instability can develop even with repulsive interactions between fermions, a number of important results, exact in the weak-coupling limit ($U\rightarrow0$), have been obtained by perturbative approaches. Baranov and Kagan \cite{baranov1992, baranov1992superconductivity, kagan1989} studied the Hubbard model in the dilute limit ($n\rightarrow0$) by 2nd-order perturbation theory. This work has been extended to the 3rd order by Chubukov and Lu \cite{chubukov1992, chubukov1993kohn} and later by Fukazawa et al. \cite{fukazawa2002}, which allowed, in particular, to obtain the boundary between different superfluid phases in the limit $n\rightarrow0$, $U\rightarrow0$. The first week-coupling phase diagram in the $n-t^{\prime}$ plane for the range of parameters $0\le t^{\prime}\le 0.5$ and $0.25 \le n \le 0.75$ was obtained by Hlubina \cite{hlubina1999phase} and the the effective coupling strengths for lines of $t^{\prime} = 0$ and $t^{\prime} = 0.3$ and $0 < n < 2$ were analysed by Raghu et al. \cite{raghu2010superconductivity} (although, as we discuss below, with algebraic mistakes that are critical for final conclusions). Of special interest is the interplay of various ordered phases when the Fermi surface is tuned to the Van Hove singularity, or in the vicinity thereof. This competition of instabilities has been inspected mainly by different renormalisation group techniques \cite{salmhofer1998, binz2003, salmhofer2001, honerkamp2001a, reiss2007, zhai2009, shankar1994, polchinski1992} at weak-coupling \cite{dzyaloshinskii1987superconducting, schulz1987, lederer1987, scalapino1986, zanchi1996superconducting, zheleznyak1997, zanchi2000weakly, katanin2003, neumayr2003, khavkine2004, yamase2005, kee2005} as well as in the strong coupling regime \cite{halboth2000, honerkamp2001, honerkamp2002, hankevych2002, hankevych2003, hankevych2003a, husemann2009, chen2015, eberlein2014}.

Very recently, the phase diagram of the Hubbard model in a wide range of parameters was studied within the random phase approximation \cite{romer2015pairing}. The approach however assumed the effective coupling in the Cooper channel $\lambda$ to be fixed whilst the value of $U$ was adjusted accordingly, so that the resulting phase diagram cannot be related to results in the $U\rightarrow0$ limit. A number of previous works applied the weak-coupling approach to the Hubbard model but evaluated the observables at strong interactions \cite{luther1994interacting, fjaerestad1999correlation, balents1996weak, lin1998exact, dimov2008competing}. Although meant to provide insight into the physics of strong correlations, such results are \textit{a priori} uncontrolled and typically deviate significantly, even qualitatively, from the (numerically) exact solution in the correlated regime whenever the latter is available \citep{deng2014emergent}. Accurate studies of the Hubbard model in the correlated regime have been possible by means of various Monte Carlo methods at half filling \cite{blankenbecler1981, staudt2000, kozik2013}, where the notorious fermionic sign problem is absent, and, more recently, with the development of advanced numerical technique, at non-zero doping values \cite{gull2013, staar2014, deng2014emergent, zheng2015, leblanc2015, meng2015}. Nonetheless, achieving full control over systematic errors in numerical studies of the doped Hubbard model in the correlated regime is still a very difficult problem \cite{leblanc2015} and a reliable phase diagram at non-zero values of $U$ is currently available only in a very limited region of the parameter space \cite{deng2014emergent}. In the context of ongoing development and testing of new numeric techniques for the Hubbard model at strong correlations, accurate results in limiting cases are indispensable. 

However, even in the weak-coupling limit $U\rightarrow0$ a complete phase diagram in the full range of filling factors $0 < n < 2$ and most relevant next-nearest-neighbour hopping amplitudes $0\le t^{\prime}\le 0.5$ is still missing. Moreover, recent results in this parameter range \cite{hlubina1999phase, raghu2010superconductivity} are in conflict with each other. A detailed analysis of the nodal structure of the Cooper-pairing order parameter for each symmetry sector in the phase diagram has not yet been carried out. Furthermore, in the context of high-temperature superconductivity, it would be of significant importance to identify regions of the phase diagram where the effective coupling strength is highest. Our paper is aimed at addressing all these issues.

We report the exact in the limit $U\rightarrow0$ ground-state phase diagram of the Hubbard model on the square lattice in the range of $0\le t^{\prime}\le 0.5$ and $0 < n < 2$. Our method consists of semi-analytical treatment of the weak Cooper instability developing in the Landau Fermi liquid (FL) at temperatures much (exponentially) smaller than the Fermi energy $E_F$. We identify twelve different superconducting phases, differentiated by the number of nodes of the superfluid order parameter, with every allowed symmetry of the order parameter represented, and study how the shape of the order parameter transforms across the boundaries between the phases in the parameter space. We perform an analysis of the effective coupling strength and identify regions of the parameter space where high-$T_c$ superconductivity might be expected at higher values of the coupling $U$. Our results fix errors in and reconcile previous studies as well as provide more detail on the structure of the order parameter in a wider range of parameters, thereby serving as solid grounds for benchmarking of new non-perturbative methods. Since obtaining controlled numeric results at essentially non-zero values of $U$ is extremely computationally expensive, our work provides a valuable guide for such studies in the search for high-$T_c$ superconductivity in the Hubbard model.   

The paper is organised as follows: In Section~\ref{subsec:perturbative-treatment} we review the method for obtaining the phase diagram by tracing the development of instability in each particular channel.  Section~\ref{subsec:group-theory} presents a brief overview of the symmetry adapted basis states on the square lattice. Section~\ref{subsec:VH} addresses competition between magnetic and superconducting instabilities along the line in the $(t^{\prime}, n)$ plane where the Van Hove singularity is at the Fermi surface. We present the obtained phase diagram in Section~\ref{subsec:phase-diagram}, and discuss the behaviour of the effective coupling strength in the Cooper channel, which controls the superfluid $T_c$, in Section~\ref{subsec:coupling}. In Section~\ref{subsec:previous-work} we compare our results to previous work, while Section~\ref{sec:conclusions} gives general concluding remarks.


\section{Method}


\subsection{Perturbative treatment of the Fermi liquid} \label{subsec:perturbative-treatment}

Our derivations follow the standard perturbative approach, adopted, e.g., in Refs.~\onlinecite{hlubina1999phase,raghu2010superconductivity}. 

The dispersion relation on the square lattice reads ($k_x$ and $k_y$ are the momentum components of $\mathbf{k}$)
\begin{align}
\epsilon(\mathbf{k}) = -2 t \left(\cos k_x + \cos k_y \right) + 4t^{\prime}\cos k_x \cos k_y .
\end{align}
The Green's function $G(\mathbf{k}, \xi)$ can be obtained from the Dyson's equation
\begin{align}
	G(\mathbf{k}, \xi_n) &= \frac{1}{\imath \xi_n -\epsilon(\mathbf{k}) + \mu -\Sigma(\mathbf{k}, \xi_n)} 	\label{G1}
\end{align}
where $\mu$ is the chemical potential, $\xi_n = (2n+1)\pi /\beta$ are the Matsubara frequencies and $\Sigma(\mathbf{k}, \xi_n)$ is the self-energy (in the following we adopt the units of the hopping amplitude $t$).

In the weak-coupling limit at sufficiently low temperatures $T \ll E_F$ the system is a Fermi liquid with a well-defined Fermi surface. The quasiparticle Green's function in the vicinity of the Fermi surface $|\xi|\ll E_F$ and $|k-k_F(\hat{k})|\ll k_F(\hat{k})$ takes on the form:
\begin{align}
G(\mathbf{k},\xi) \equiv \frac{z(\hat{k})}{i\xi - \mathbf{v}_F(\hat{k})\times[\mathbf{k}-\mathbf{k}_F(\hat{k})]}
\label{G2}
\end{align}
Here the Fermi surface is parametrised in terms of the Fermi momentum $\mathbf{k}_F(\hat{k})$ in the direction $\hat{k}$ of the vector $\mathbf{k}$. Comparing Eq.~\ref{G1} and Eq.~\ref{G2} we obtain the Fermi velocity 
 $\mathbf{v}_F(\hat{k}) = z(\hat{k}) \nabla_{\mathbf{k_F}} (\epsilon(\mathbf{k_F}) + \Sigma_R(\mathbf{k_F}, \xi_n))$ and the quasiparticle residue $z(\hat{k}) = (1-\lim_{\xi_n \rightarrow 0}\frac{\Sigma_I(\mathbf{k_F}, \xi_n)}{\xi_n})^{-1}$. 
 
As the temperature is lowered further, the development of the Cooper instability is marked by divergence of the pairing susceptibility at the critical temperature $T_c$, which is exponentially smaller than $E_F$. 
 \begin{figure}[!ht]
  \centering
  \begin{picture}(50,30)
  \put(-85,0){\includegraphics[height=1cm]{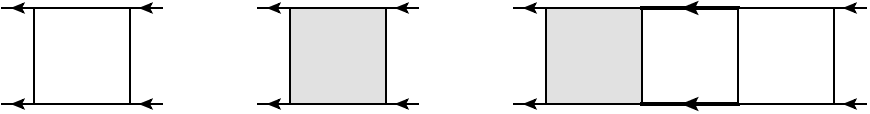}}
\put(-60,-5){$-p_2$}\put(-91,-5){$-p_1$}\put(-52,32){$p_2$}\put(-83,32){$p_1$}
\put(-72,11){$\hat{F}^{\text{pp}}$}
\put(-36,12){$=$}
\put(4,-5){$-p_2$}\put(-28,-5){$-p_1$}\put(12,32){$p_2$}\put(-20,32){$p_1$}
\put(-8,11){$\hat{\Gamma}^{\text{pp}}$}
\put(28,12){$+$}
\put(116,-5){$-p_2$}\put(36,-5){$-p_1$}\put(78,-5){$-p_3$}\put(124,32){$p_2$}\put(44,32){$p_1$}\put(86,32){$p_3$}
\put(57,11){$\hat{\Gamma}^{\text{pp}}$}
\put(104,11){$\hat{F}^{\text{pp}}$}
\end{picture}
\caption{The Bethe-Salpeter equation for $\Gamma^{\text{pp}}$ with $p_i \equiv \left( \xi_i, \mathbf{k}_i \right)$. Summation over $\xi_i$ and integration over $\mathbf{k}_i$ is assumed.}
\label{bse}
\end{figure}
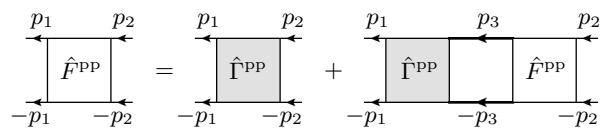
The instability is due to weak attraction between fermions, which in our case is an emergent low-energy many-body property. Mathematically, the effective interaction is described by the irreducible in the particle-particle channel four-point vertex $\Gamma^{\text{pp}}$, which in general is a sum of all possible four-point diagrams that cannot be split into disconnected pieces by cutting two particle lines. The Cooper pairing susceptibility is proportional to the full effective vertex $F^{\text{pp}}$, which diverges at $T_c$ and is related to $\Gamma^{\text{pp}}$ via the Bethe-Salpeter equation shown diagrammatically in Fig.~\ref{bse}. From the Bethe-Salpeter equation we see that the smallness of the attractive part of $\Gamma^{\text{pp}}$ is a natural condition preventing $F^{\text{pp}}$ from dramatic growth at $T \ll E_F$. Indeed, in the FL regime, the leading contribution to the integral over $\mathbf{k}_3$ in the second term on the r.h.s. of Fig.~\ref{bse} comes from  the close vicinity to the Fermi surface, 
\begin{align}
\int \operatorname{d} \mathbf{k}_3 \sum_{\xi_3} G(p_3) G(-p_3) \approx \ln\frac{c E_f}{T} \int \frac{\operatorname{d} s}{2 \pi} \frac{z^2(\hat{k})}{2 \pi v_F},
\end{align}
where $\operatorname{d} s$ is the Fermi surface element. Only the finite temperature (i.e., discreteness of Matsubara frequency $\xi_3$) prevents the integral in the r.h.s. from logarithmic divergence. With logarithmic accuracy at
$T\ll E_F$ , we have
\begin{align}
F^{\text{pp}}_{\hat{k}_1,\hat{k}_2} \approx \Gamma^{\text{pp}}_{\hat{k}_1,\hat{k}_2}+ \ln\frac{c E_f}{T} \int \Gamma^{\text{pp}}_{\hat{k}_1,\hat{k}_3} Q_{\hat{k}_3}  F^{\text{pp}}_{\hat{k}_3,\hat{k}_2} \operatorname{d} \hat{k}_3
\label{systematicerror}
\end{align}
where $F_{\hat{k}_1,\hat{k}_2}$ and $\Gamma_{\hat{k}_1,\hat{k}_2}$ are $F^{\text{pp}}$ and $\Gamma^{\text{pp}}$ at vanishing frequencies projected to the Fermi surface:
\begin{align}
F^{\text{pp}}_{\hat{k}_1,\hat{k}_2} \equiv F^{\text{pp}} (\mathbf{k}_1 = \mathbf{k}_F (\hat{k}_1), \xi_1 \rightarrow 0;\mathbf{k}_2 = \mathbf{k}_F (\hat{k}_2), \xi_2 \rightarrow 0)
\end{align}
and $Q_{\hat{k}}$ is the product of $z^2 (\hat{k})$ and the density of states at the $k$-point on the Fermi surface.
\begin{align}
Q_{\hat{k}} = \frac{k_F (\hat{k}) z^2 (\hat{k})}{2\pi \, v_F (\hat{k})}
\end{align}
Switching to matrix notations $F^{\text{pp}}_{\hat{k}_1,\hat{k}_2} \rightarrow \hat{F}^{\text{pp}}$, $\Gamma^{\text{pp}}_{\hat{k}_1,\hat{k}_2} \rightarrow \hat{\Gamma}^{\text{pp}}$, $Q_{\hat{k}_2} \rightarrow \hat{Q}$ we find
\begin{align}
\hat{F}^{\text{pp}} \approx \left[ 1- \ln (\frac{E_F}{T}) \hat{\Gamma}^{\text{pp}} \hat{Q}\right]^{-1} \hat{\Gamma}^{\text{pp}},
\label{bsp}
\end{align}
implying that $\hat{F}^{\text{pp}}$, and thus the static response function in the Cooper channel, diverges at the critical temperature 
\begin{align}
T_c = c \, E_F\, e^{-1/\lambda}
\end{align}
where $\lambda$ is the largest positive eigenvalue of $\hat{\Gamma}^{\text{pp}} \hat{Q}$. 
Solving the problem with logarithmic accuracy (which is guaranteed in the $U \to 0$ limit due to $\lambda \to 0$ and the corresponding exponential smallness of $T_c$) amounts to finding the eigenvalues/eigenvectors of a real symmetric matrix
\begin{align}
T_{\hat{k}_1,\hat{k}_2} \psi_{\hat{k}_2} = \lambda \,  \psi_{\hat{k}_1}, \quad T_{\hat{k}_1,\hat{k}_2}  = Q^{\frac{1}{2}}_{\hat{k}_1} \Gamma_{\hat{k}_1,\hat{k}_2} Q^{\frac{1}{2}}_{\hat{k}_2} \label{eigenvalue_problem}
\end{align}
where the eigenvector $\psi_{\hat{k}}$ is the wave function of the Cooper pair in the momentum representation. 

The effective vertex $\hat{\Gamma}^{\text{pp}}$ can be computed as a diagrammatic expansion in the bare coupling $U$. In this expansion the first order diagram is a negative constant $-U$, which by itself can never lead to a diverging denominator in Eq.~\ref{bsp}. 
 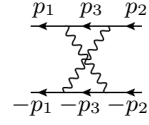
\begin{figure}[!ht]
  \centering
  \begin{picture}(50,30)
\put(-5,0){\includegraphics[height=1cm]{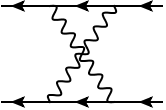}}
\put(24,-5){$-p_2$}\put(-12,-5){$-p_1$}\put(6,-5){$-p_3$}
\put(31,32){$p_2$}\put(-4,32){$p_1$}\put(14,32){$p_3$}
\end{picture}
\caption{The second-order diagram contributing to $\hat{\Gamma}^{\text{pp}}$. The wavy lines are the interaction vertices $U$, the straight lines with arrows are the non-interacting propagators $G_0$. Integration over internal momenta is assumed.  }
\label{second}
\end{figure}
The first non-vanishing contribution to Cooper pairing comes from the second order in $U$ diagram, shown in Fig.~\ref{second}, which features non-trivial momentum dependence giving rise to positive eigenvalues of the matrix $T_{\hat{k}_1,\hat{k}_2} \psi_{\hat{k}_2}$. All the diagrams beyond second order are vanishing in the limit $U \to 0$, and can be neglected. With the same accuracy, the propagator lines in the diagram in Fig.~\ref{second} are given by the bare non-interacting Green's function $G_0$, i.e. the self-energy contribution in Eq.~\ref{G1} can be neglected giving
\begin{align}
G_0(\mathbf{k}, \xi_n) &= \frac{1}{\imath \xi_n -\epsilon(\mathbf{k}) + \mu}. \label{G_0}
\end{align}
Correspondingly, in Eq.~\ref{G2}, the quasiparticle residue $z(\hat{k})=1$. Thus, the diagram Fig.~\ref{second} is given by
\begin{align}
\Gamma_{\hat{k}_1,\hat{k}_2} \approx {\chi^{\text{ph}}}_{\hat{k}_1,\hat{k}_2} = -\int \frac{\operatorname{d}^d q}{(2\pi)^d} \frac{  \nu(\epsilon(k+q)) - \nu(\epsilon(k))}{\epsilon(k+q)-\epsilon(k)}, \label{Gamma_o2}
\end{align}
where $\nu(\epsilon)=[1+\exp((\epsilon-\mu)/T)]^{-1}$ is the Fermi-Dirac distribution function. In two dimensions it is convenient to parametrise $\hat{k}$ with the polar angle $\theta$ and to write the eigenvalue/eigenvector problem explicitly as 
\begin{align}
\int^{2\pi}_0 \frac{\operatorname{d} \theta_2}{2\pi} T_{\theta_1,\theta_2} \psi_{\theta_2} = g \,\psi_{\theta_1}, \quad T_{\theta_1,\theta_2} = Q^{\frac{1}{2}}_{\theta_1} \Gamma_{\theta_1,\theta_2} Q^{\frac{1}{2}}_{\theta_2}.  \label{eigenvalue_problem_2}
\end{align}
Note that this parametrisation only works for connected Fermi surfaces, which is the case for all of parameter space in our model for values $t^{\prime} \le 0.5$. 

We employ the following protocol to obtain the ground state phase diagram:
\begin{enumerate}
	\item  For a given set of parameters $(t^{\prime},n)$, the Fermi surface is found as the pole of $G_0$, Eg.~\ref{G_0}, in the limit $T \to 0$, which gives $v_F(\hat{k)}$ and $k_F(\hat{k})$.  
	\item	The matrix $\mathit{\Gamma}_{\theta_1,\theta_2}$ is computed using Eq.~\ref{Gamma_o2} by Monte Carlo numerical integration. 
	\item	The eigenvalue problem, Eq.~\ref{eigenvalue_problem_2}, is solved in the basis given by the point symmetry group (as explained below). 
	\item 	 The largest eigenvalue  $\lambda$ and the corresponding eigenvector determines the superfluid ground state realised at the given set of parameters. 
\end{enumerate}

 \begin{figure}[!ht]
\begin{center}
\begin{tabular}{ |c||c|c|c|c|cc| } 
 \hline 
 m & $A_1$ & $A_2$ & $B_1$ & $B_2$ & $E_1 (a)$&$ E_1 (b)$\\ 
 \hline \hline
0 &$s^{(0)}$ & $g^{(8)}$ &$d_{xy}^{(4)}$ & $d_{x^2-y^2}^{(4)}$ & $p_x^{(2)}$ & $p_y^{(2)}$\\ 
 & \includegraphics[width=1.2cm]{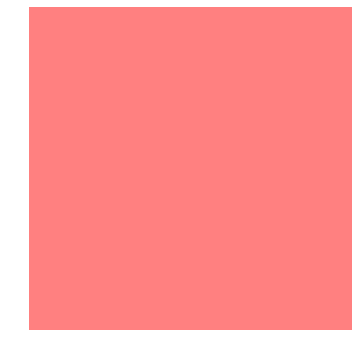}   & \includegraphics[width=1.2cm]{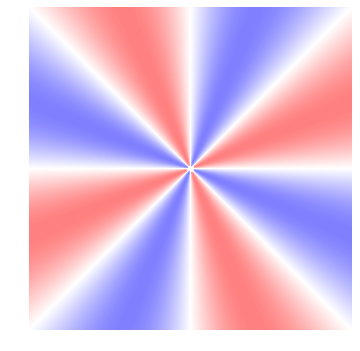}  & \includegraphics[width=1.2cm]{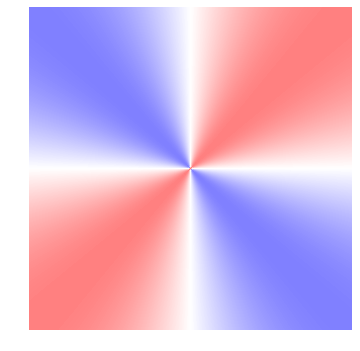}  & \includegraphics[width=1.2cm]{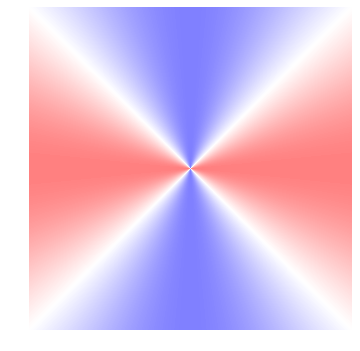}  &  \includegraphics[width=1.2cm]{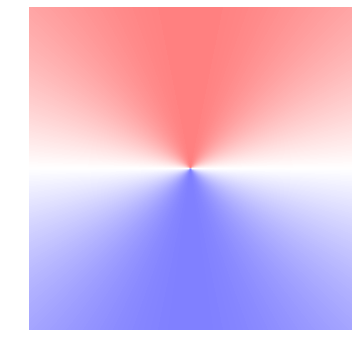} & \includegraphics[width=1.2cm]{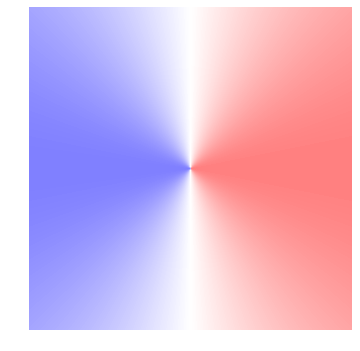} \\ \hline
 1 &$s^{(8)}$ & $g^{(16)}$ &$d_{xy}^{(12)}$ & $d_{x^2-y^2}^{(12)}$ & $p_x^{(6)}$ & $p_y^{(6)}$\\ 
  & \includegraphics[width=1.2cm]{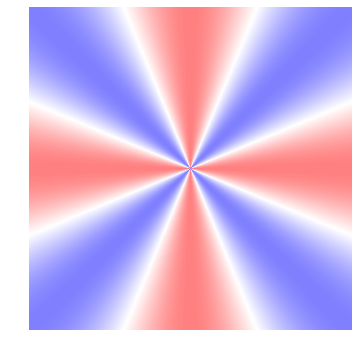}  & \includegraphics[width=1.2cm]{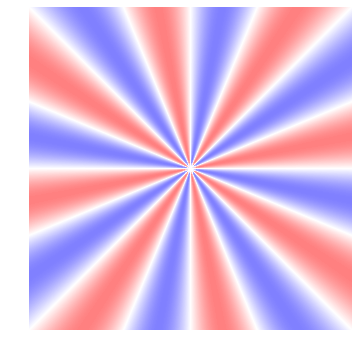}  & \includegraphics[width=1.2cm]{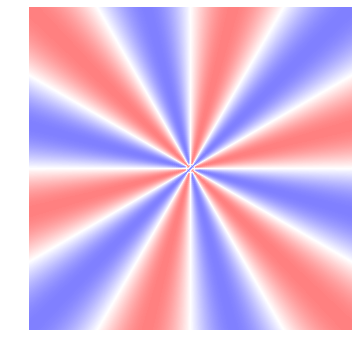}  &  \includegraphics[width=1.2cm]{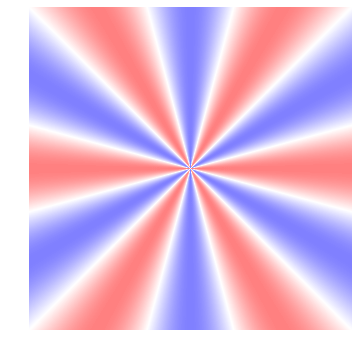} & \includegraphics[width=1.2cm]{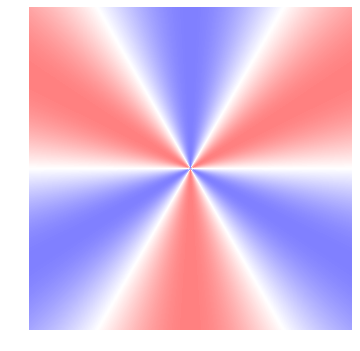} & \includegraphics[width=1.2cm]{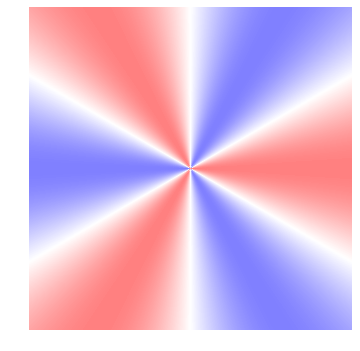}\\ \hline
2 &$s^{(16)}$ & $g^{(24)}$ &$d_{xy}^{(20)}$ & $d_{x^2-y^2}^{(20)}$ & $p_x^{(10)}$ & $p_y^{(10)}$\\ 
  & \includegraphics[width=1.2cm]{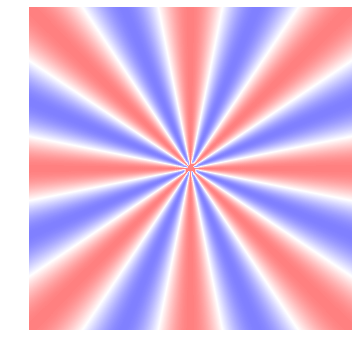}  & \includegraphics[width=1.2cm]{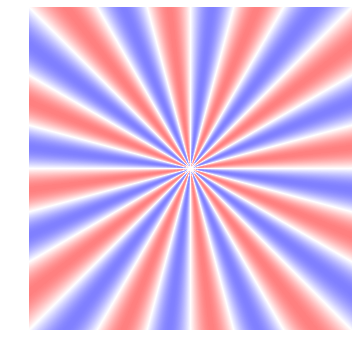}  & \includegraphics[width=1.2cm]{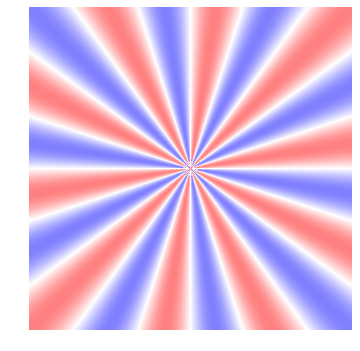}  & \includegraphics[width=1.2cm]{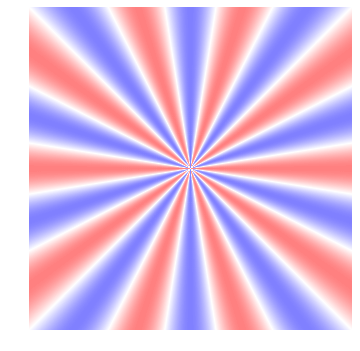}  & \includegraphics[width=1.2cm]{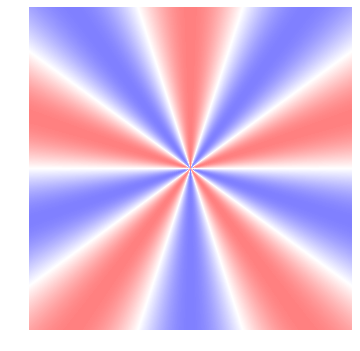} &  \includegraphics[width=1.2cm]{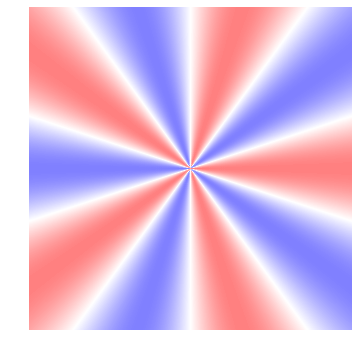}\\ 
 \hline
\end{tabular}
\end{center}
\vspace{-0.5cm}
\caption{The form of basis functions in the Brillouin zone (red, blue and white colours correspond to positive, negative values and nodes respectively) categorised by irreducible representation and the order of harmonic $m$. (See  Eqs.~\ref{harmonics1}-\ref{harmonics2}).}
\label{wavesclasses}
\end{figure}


\subsection{Classification of basis states on the square lattice} \label{subsec:group-theory}

The symmetry operations on the square lattice form the point group $D_4$ (the two-dimensional point group corresponding to $D_{4h}$). The operations are the identity ($E$), two rotations by $\pm\pi/2$ ($C_4$) and one rotation by $\pi$ ($C_2$) in the main symmetry axis (perpendicular to the plane) as well as two rotations by $\pi$ around the horizontal/vertical in-plane axes ($C_2^{\prime}$) and two $\pi$ rotations around the diagonal in-plane axes ($C_2^{\prime \prime}$). 

The $D_4$ symmetry  dictates that the matrix $\mathit{\Gamma}_{\theta_1,\theta_2}$ splits into four independent singlet blocks, known as  $s$, $g$, $d_{xy}$ and $d_{x^2-y^2}$, which correspond to one-dimensional irreducible representations $A_{1}$, $A_{2}$, $B_{1}$, $B_{2}$, and the doubly degenerate triplet block $p$, which corresponds to the two-dimensional irreducible representation $E_1$. The $E_1$-sector further splits into $p_x$ and $p_y$ eigenvalues/eigenfunctions, which are related to each other by a $\pm \pi/2$ rotation. For each of the six sectors, the symmetry properties of the corresponding eigenvectors $\Psi(\theta)$ are readily seen from their Fourier expansions (with integer values of $m$)
\begin{align}
A_{1}:&\; \Psi_s (\theta) = \sum_{m=0}^{\infty} A_m \cos\left( 4m\, \theta \right) \label{harmonics1} \\
A_{2}:&\; \Psi_g (\theta) = \sum_{m=0}^{\infty} B_m \sin\left( \left(4m+4\right) \theta \right) \\
B_{1}:&\; \Psi_{d_{xy}}  = \sum_{m=0}^{\infty} C_m \cos\left( \left(4m+2\right) \theta \right)  \\
B_{2}:&\; \Psi_{d_{x^2-y^2} } (\theta) = \sum_{m=0}^{\infty} D_m \sin\left( \left(4m+2\right) \theta \right) \\
E_{1}: \;  & \begin{cases}
          \Psi_{p_x} (\theta) = \sum_{m=0}^{\infty} E_m \cos\left( \left(2m+1\right) \theta \right) \\
          \Psi_{p_y} (\theta) = \sum_{m=0}^{\infty} E_m \sin\left( \left(2m+1\right) \theta \right) 
    \end{cases}
    \label{harmonics2}
\end{align}
The eigenfunctions $\Psi_s$ are invariant under all symmetry operations of the point group. The eigenfunctions $\Psi_g$ are invariant under $E$, $C_4$ and $C_2$, but change sign under $C_2^{\prime}$ and $C_2^{\prime \prime}$. Both $\Psi_{d_{xy}}$ and $\Psi_{d_{x^2-y^2}}$ change signs under $C_4$, while only $\Psi_{d_{xy}}$ changes sign under $C_2^{\prime}$ and only  $\Psi_{d_{x^2-y^2}}$ changes sign under $C_2^{\prime \prime}$. Finally $\Psi_{p_x}$ and $\Psi_{p_y}$ transform into each other under $C_4$ and into a linear combination thereof under all the other symmetry operations. We refer to the $m=0$ contribution to each eigenfunction in Eqs.~\ref{harmonics1}-\ref{harmonics2} as the corresponding fundamental mode and all the eigenfunctions with $m>0$ as higher harmonics. We assign a number in the superscript to each eigenfunction, which signifies the amount of zeros of the function. An example of the fundamental mode and the first two higher harmonics projected onto the Brillouin zone of the square lattice is given in Fig.~\ref{wavesclasses}. It must be noted that identifying the largest coefficient in the expansion in Eqs.~\ref{harmonics1}-\ref{harmonics2} for each eigenfunction is not always sufficient to classify the eigenfunction in terms of the number of nodes it features since the sub-leading components can have a significant net contribution that can change the nodal structure. We therefore classify each state in the phase diagram by explicitly counting the number of zeros in the eigenfunction that corresponds to the largest eigenvalue.

 \begin{figure}[h!]
  \centering
\hspace{-0.5cm}\includegraphics[width=8cm]{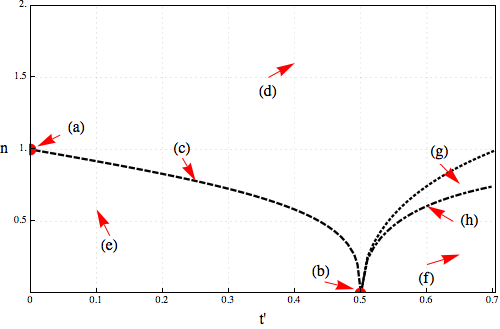} \\
\vspace{-0.3cm}
 \begin{center}
\hspace{-0.2cm}
\begin{tabular}{ cccccccc } 
(a) & (b) & (c) & (d) & (e) & (f) & (g) & (h) \\ 
\includegraphics[width=0.9cm]{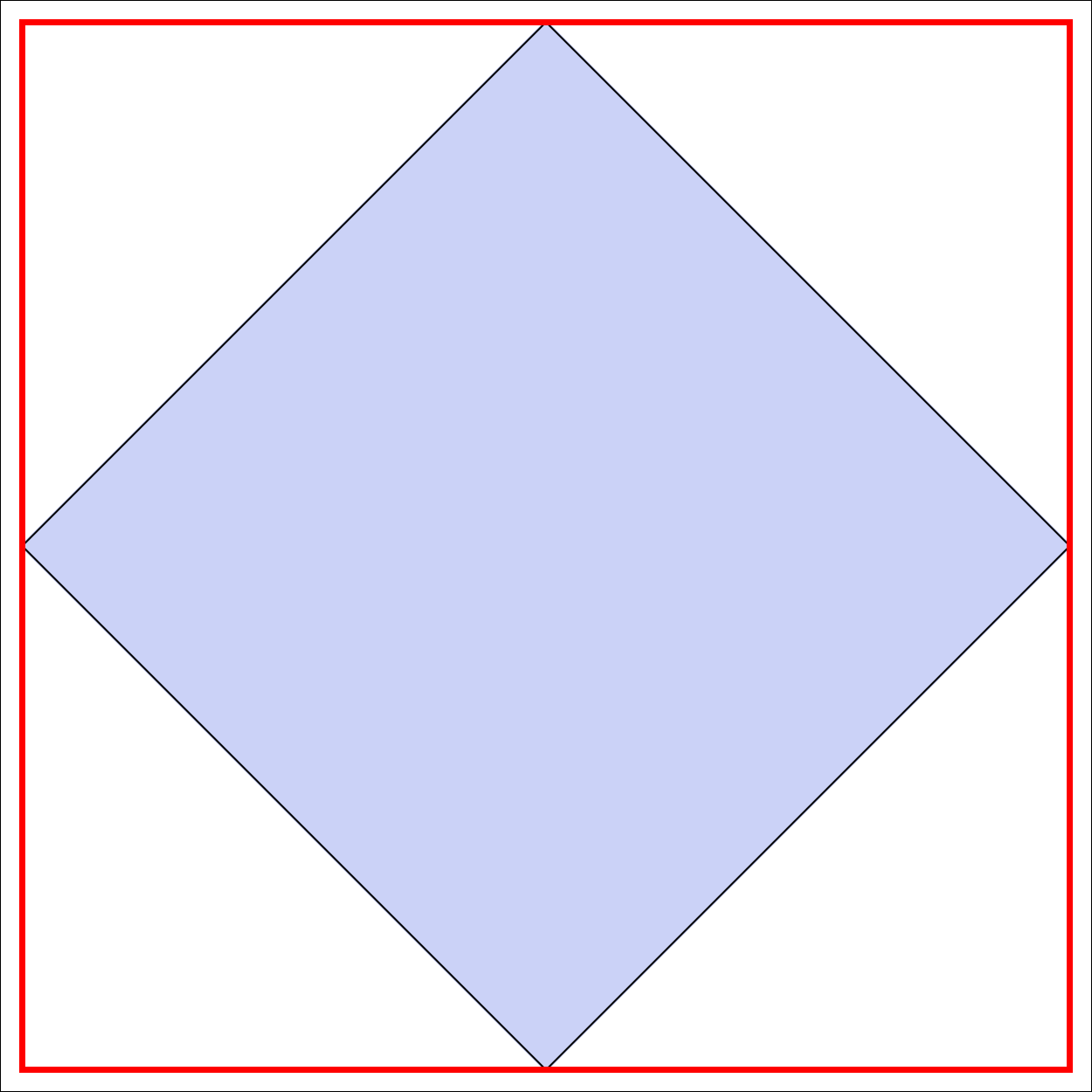} &
\includegraphics[width=0.9cm]{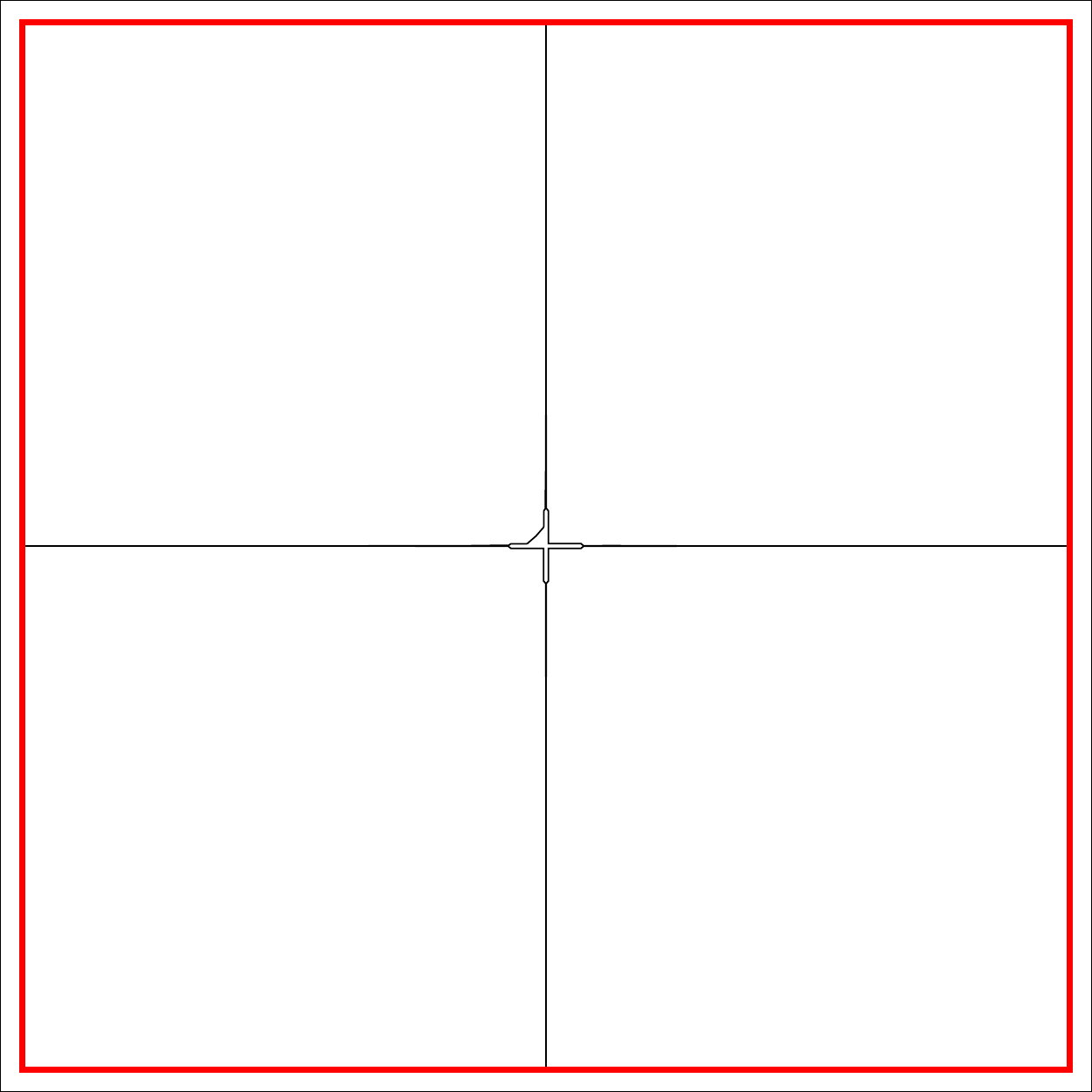} &
\includegraphics[width=0.9cm]{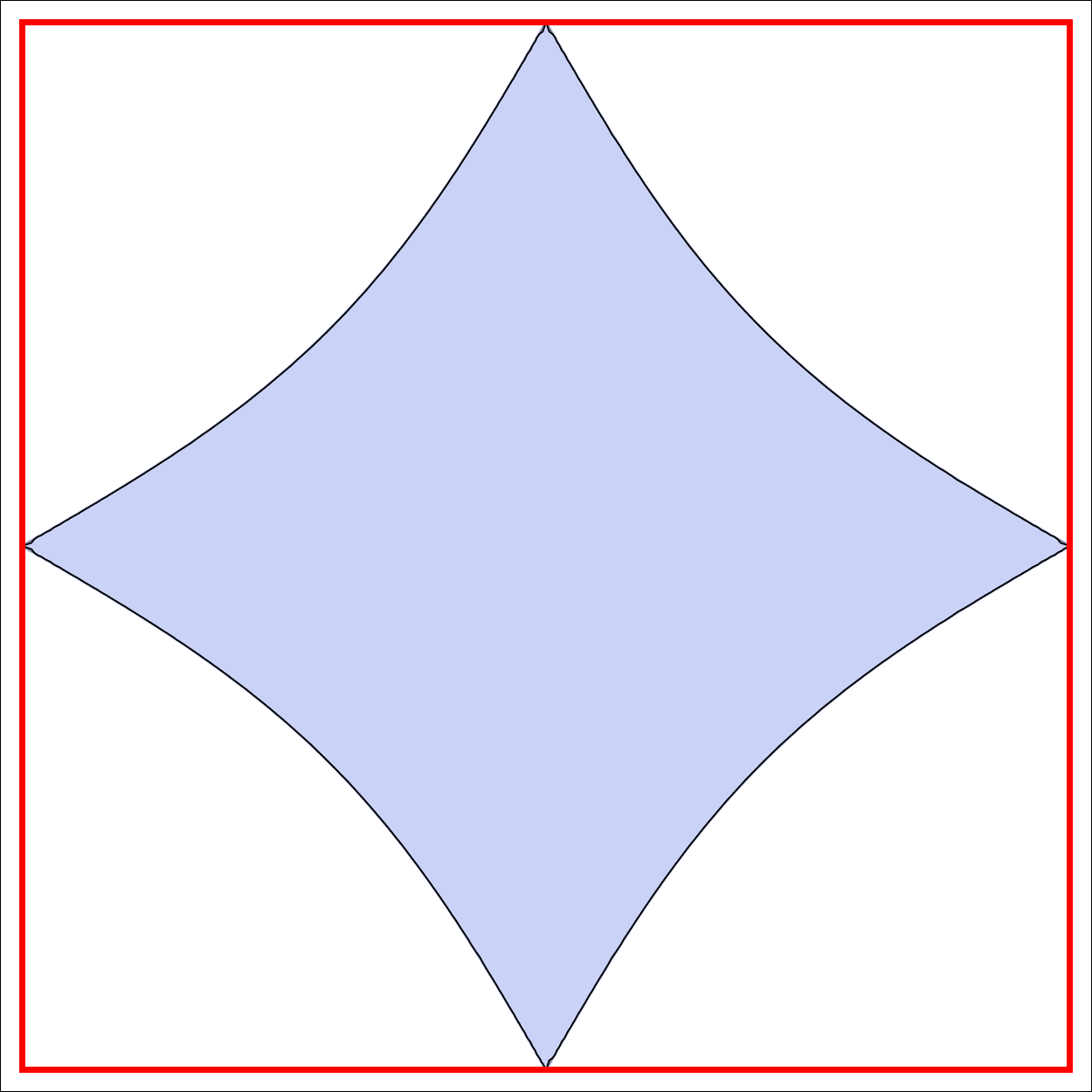} &
\includegraphics[width=0.9cm]{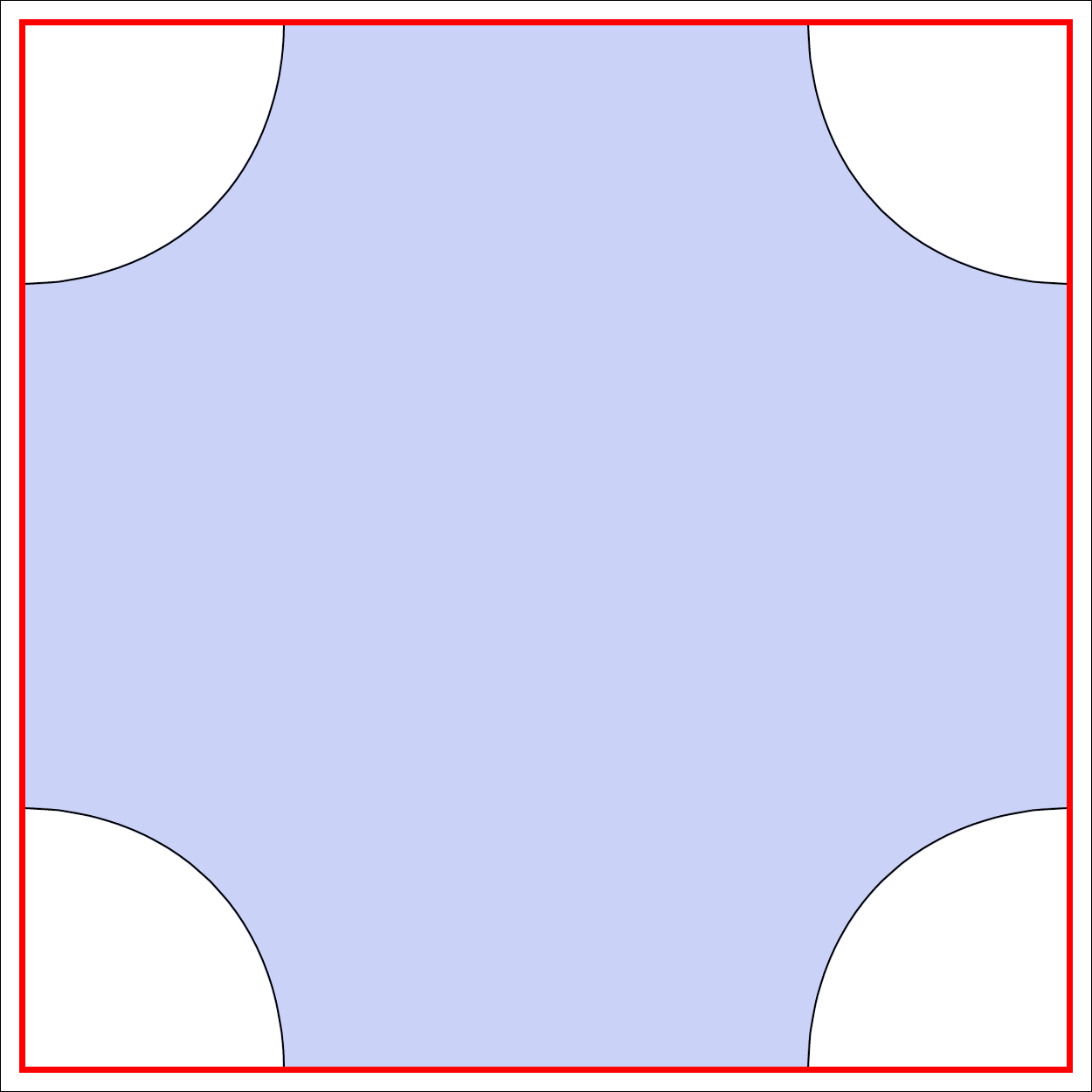} & 
\includegraphics[width=0.9cm]{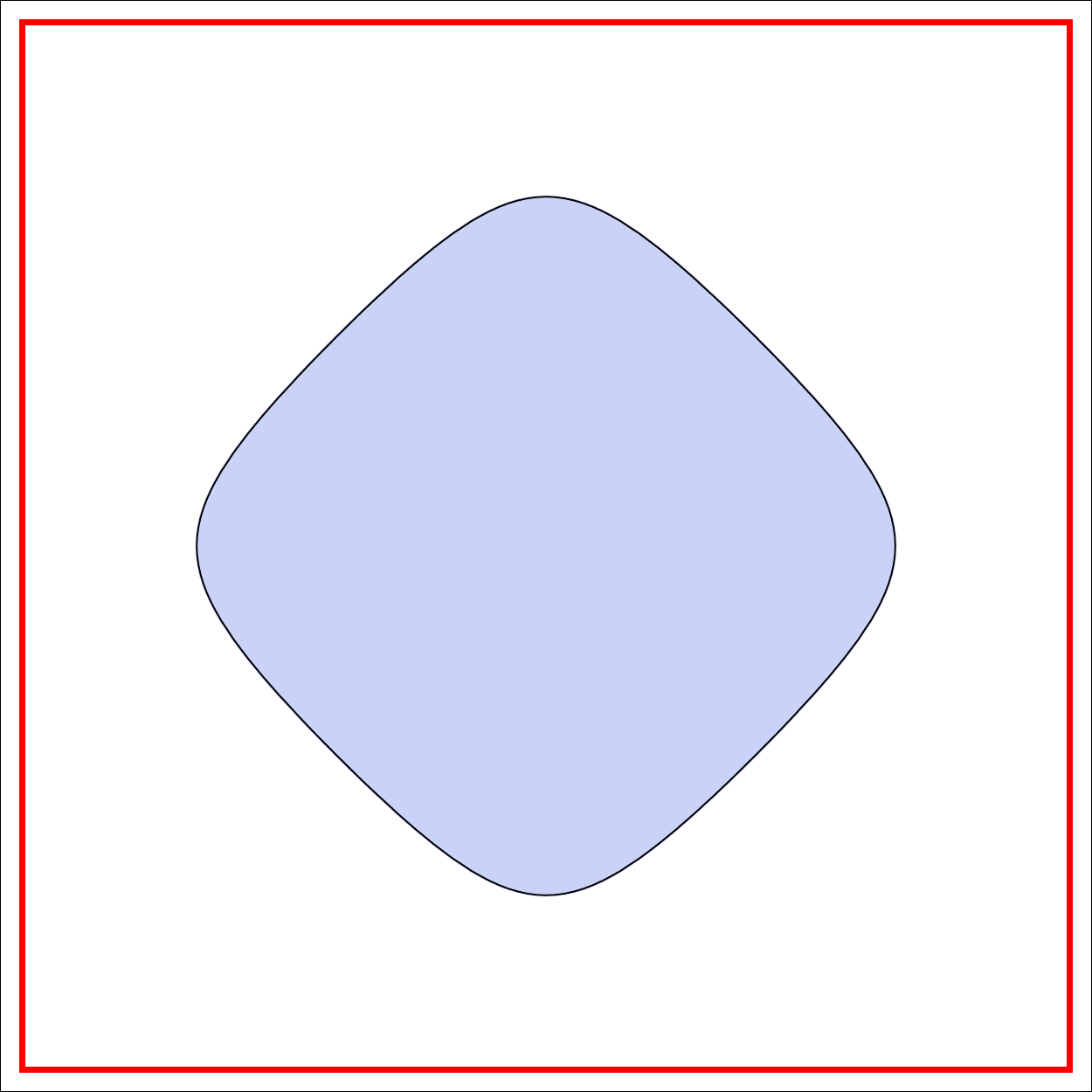} &
\includegraphics[width=0.9cm]{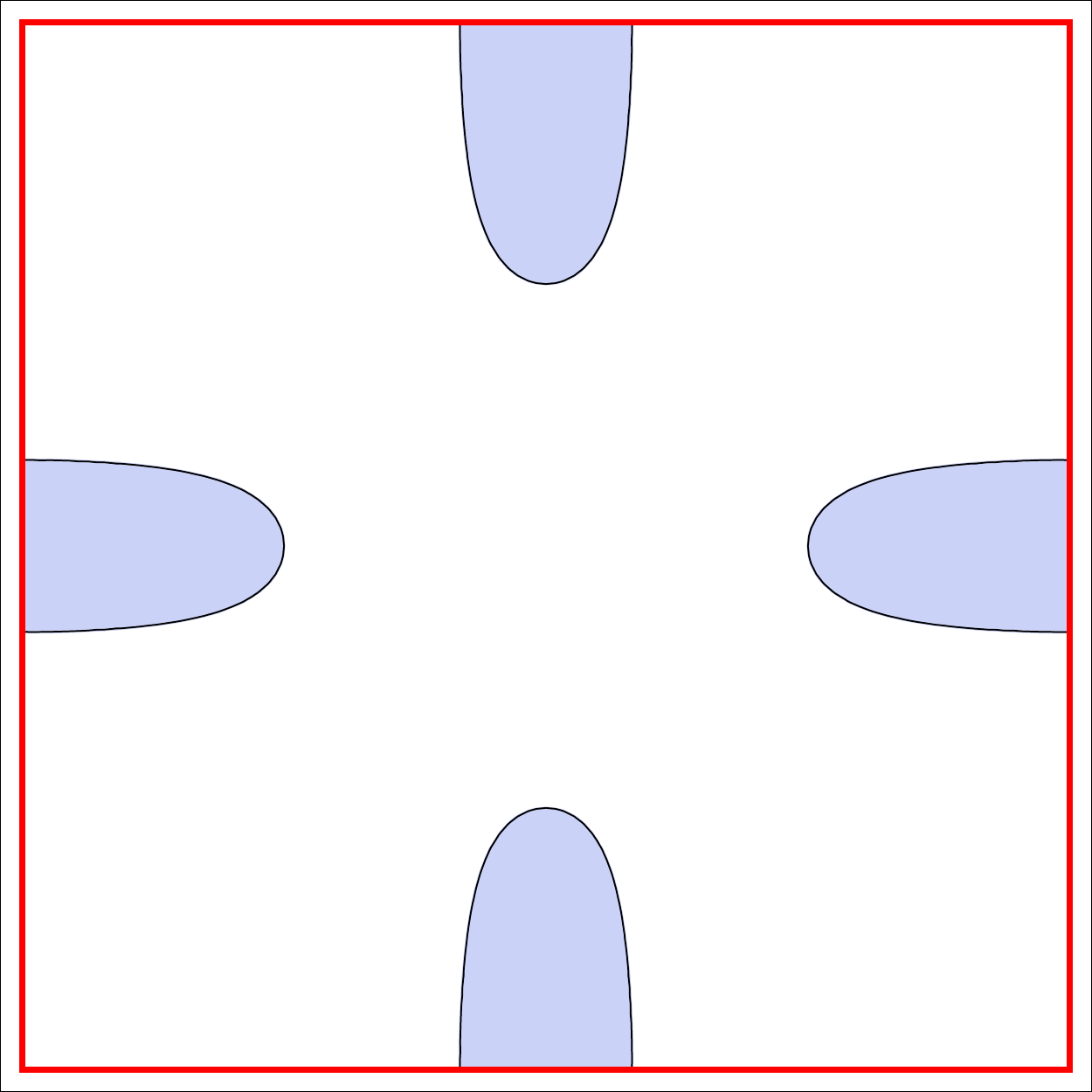} &
\includegraphics[width=0.9cm]{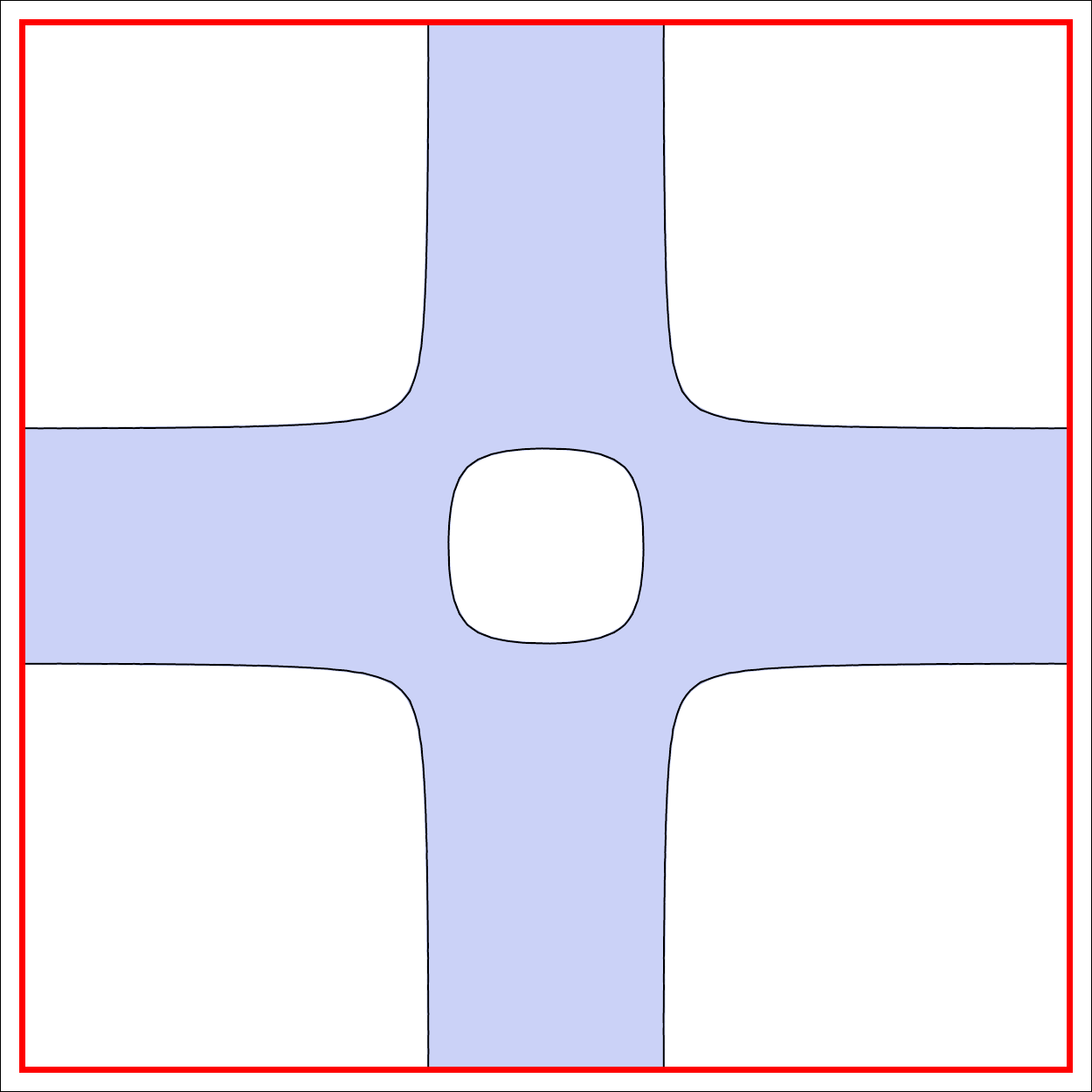} &
\includegraphics[width=0.9cm]{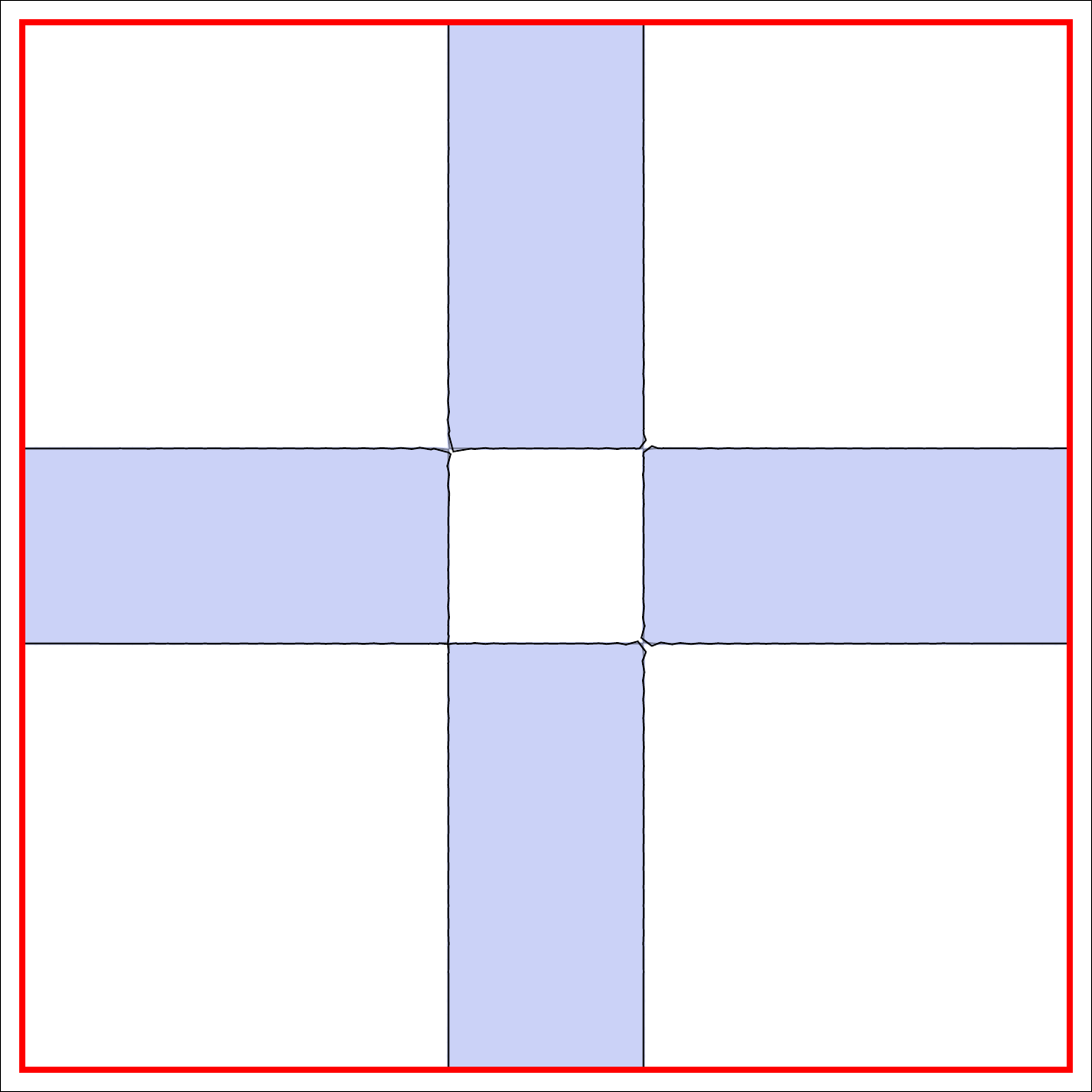} 
\end{tabular}
\end{center}
\vspace{-0.5cm}
\caption{Special lines in the parameter space $0 \le n \le 2$, $0 \le t^{\prime}  \le 0.7$: the Van Hove line (dashed), line of nesting (dot-dashed), and the dotted line beneath which a Fermi pocket appears down to the line of nesting.  The position and characteristic Fermi surfaces in the first Brillouin zone (of size $k_x, k_y \in (-\pi, \pi)$) are shown for: (a) the antiferromagnetic point \{$n=1$, $t^{\prime}=0$\}, (b) the ferromagnetic point \{$n=0$, $t^{\prime}=0.5$\}, (c) an arbitrary point on the Van Hove line \{$n=0.78$, $t^{\prime}=0.25$\}, (d) above the Van Hove line \{$n=1.60$, $t^{\prime}=0.4$\}, (e) below the Van Hove line \{$n=0.58$, $t^{\prime}=0.1$\}, (f) below the second nested line \{$n=0.27$, $t^{\prime}=0.65$\}, (g) above the line of nesting, in a region where a Fermi pocket exists around $(k_x,k_y) = (0,0)$ \{$n=0.76$, $t^{\prime}=0.65$\}, (h) on the line of nesting \{$n=0.61$, $t^{\prime}=0.6$\}.}
\label{FS}
\end{figure}


\subsection{Instabilities along the Van Hove singularity line} \label{subsec:VH}

In the weak-coupling limit, the superfluid phase is realised in the ground state in the whole parameter range with the exception of two points. Both of them lie on the line in the ($t^{\prime}, n)$ plane where the density of states at the Fermi surface diverges due to the Van Hove singularity. This line (referred to as Van Hove line) is defined by the condition $\mu = 4 t^{\prime}$ (dashed line in Fig.~\ref{FS}). At $t^{\prime}=0$ and $n=1$, due to nesting of the Fermi surface with the momentum $\mathbf{Q}_{AFM} = (\pi,\pi)$, the spin ordering instability dominates, and the ground state is an antiferromagnet. At $t^{\prime}=0.5, \, n=0$ the ferromagnetic instability with the nesting vector $\mathbf{Q}_{FM}=(0,0)$ is leading. Generally, along the line, the particle-hole susceptibility $\chi^{\text{ph}}$ diverges double logarithmically\cite{hlubina1997ferromagnetism} at momentum transfer $\mathbf{Q}=(\pi,\pi)$ and logarithmically at $\mathbf{q}=(0,0)$ as:
\begin{align}
\chi^{\text{ph}}(\mathbf{q})  &\sim \left(\frac{1}{2\pi^2} \right) \frac{1}{\sqrt{1-4{t^{\prime}}^2}} \ln\left(\frac{1}{\Lambda} \right) \\
\chi^{\text{ph}}(\mathbf{Q}) &\sim \left(\frac{1}{2\pi^2} \right) \ln\left( 1+ \sqrt{1-4{t^{\prime}}^2} \right) \ln\left(\frac{1}{\Lambda} \right) 
\end{align}
where $\Lambda$ is the infrared energy cutoff. The particle-particle susceptibility also diverges at $\mathbf{q}$ and $\mathbf{Q}$:
\begin{align}
\chi^{\text{pp}}(\mathbf{q})  &\sim \left(\frac{1}{4\pi^2} \right) \frac{1}{\sqrt{1-4{t^{\prime}}^2}} \ln^2\left(\frac{1}{\Lambda} \right) \\
\chi^{\text{pp}}(\mathbf{Q}) &\sim \left(\frac{1}{2\pi^2} \right) \frac{\tan^{-1} \left(\frac{2t^{\prime}}{\sqrt{1-4{t^{\prime}}^2}}\right)}{\sqrt{1-4{t^{\prime}}^2}} \ln\left(\frac{1}{\Lambda} \right) 
\end{align}
Since both particle-particle and particle-hole channels are divergent along the Van Hove line, the magnetic and superfluid instabilities fuel each other and a simple Bethe-Salpeter analysis is insufficient. The behaviour in this regime has been extensively studied by means of RG \cite{zanchi2000weakly} and parquet approximation \cite{irkhin2001effects}. In particular, it was shown that $d_{x^2-y^2}^{(4)}$ state is dominant along the Van Hove line at small $U$. Larger values of $U$ were also addressed in these studies, but the methods are not controlled there and exact results are still due. At hopping values $t^{\prime} \ge 0.5$ another line of special interest exists, which is defined by the Van Hove singularity crossing the Fermi surface nested with the momentum $q=(k,k)$, $k=\pm \cos^{-1}(\frac{1}{2 | t^{\prime}|})$ \cite{romer2015pairing}, which we refer to as the line of nesting. The line starts from the ferromagnetic point and is given by the equation $\mu = -\frac{1}{t^{\prime}}$ (dot-dashed line in Fig.~\ref{FS}). It appears that the physics in the vicinity of $t^{\prime}=0.5$, even for $t^{\prime} \le 0.5$, is largely influenced by this line (see below). Above this line there is a finite region, bound above by the line $\mu=-4+4t^{\prime}$ (dotted line in Fig.~\ref{FS}), where the Fermi surface has a pocket around $(k_x, k_y) = (0,0)$.


\section{results}


\subsection{Phase diagram} \label{subsec:phase-diagram}

 \begin{figure}[!ht]
  \centering
\includegraphics[height=7cm]{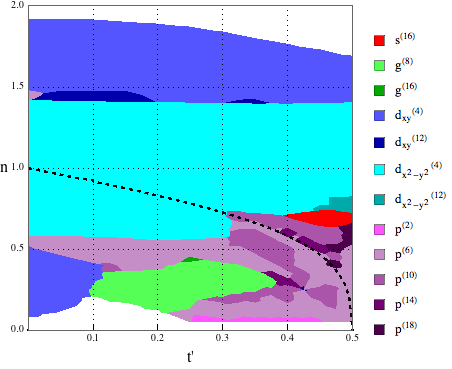} \\
\hspace{-0.7cm} \includegraphics[height=7cm]{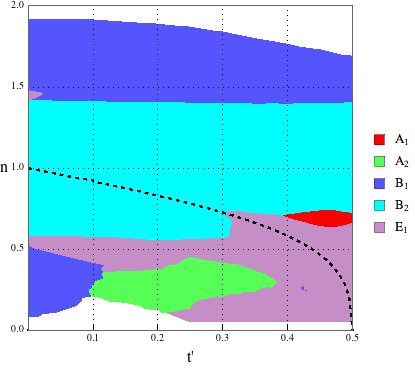}
\vspace{-0.3cm}
\caption{top: Phase diagram for the parameter range of $0<n<2$ and $0<t^{\prime}<0.5$. bottom:  same parameter range by leading irreducible representation. The presence of a Van Hove singularity is portrayed by the black dashed line.}

\label{phasediagram}
\end{figure}

 \begin{figure*}[!ht]
\begin{center}
\begin{tabular}{ ccccccc } 
\hspace{0.6cm}$ n=0.05$ &$n=0.10$ & $n=0.15$ &$n=0.20$ & $n=0.25$ & $n=0.30$ & $n=0.35$ \\ 
 \raisebox{-0.53cm}{\includegraphics[width=2.91cm]{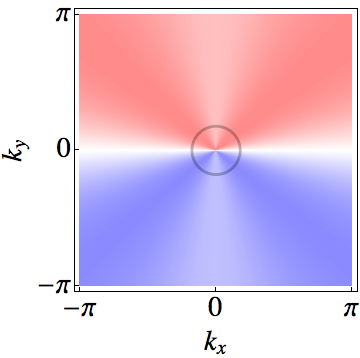}} & \includegraphics[width=2.3cm]{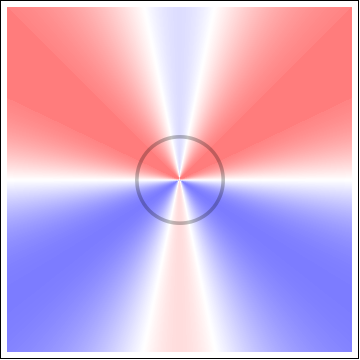}   & \includegraphics[width=2.3cm]{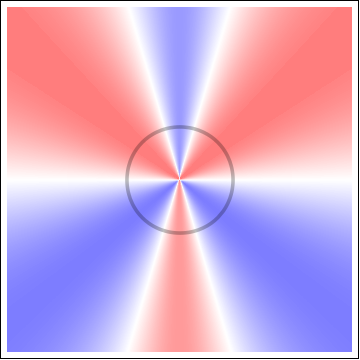}  & \includegraphics[width=2.3cm]{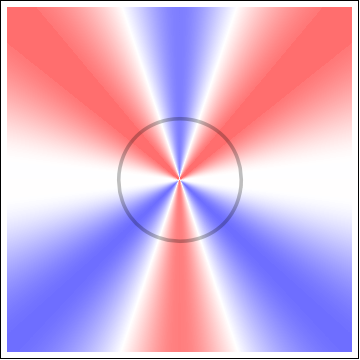}  & \includegraphics[width=2.3cm]{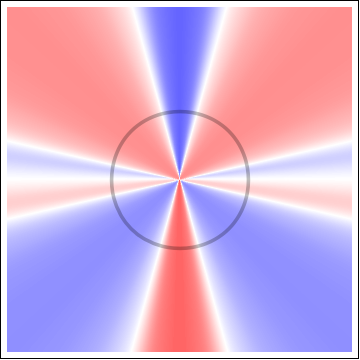}  &  \includegraphics[width=2.3cm]{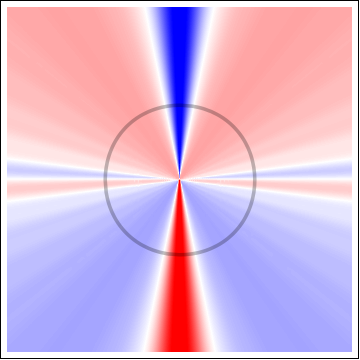} & \includegraphics[width=2.3cm]{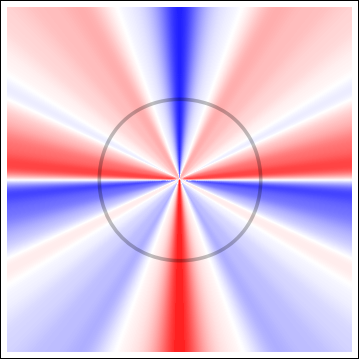} \\
 $p^{(2)}$ & $p^{(6)}$ & $p^{(6)}$ & $p^{(6)}$ & $p^{(10)}$ & $p^{(10)}$ ($g^{(8)}$) & $p^{(14)}$
\end{tabular}
\end{center}
\vspace{-0.3cm}
\caption{The transition between the first four waves of the $E_1$ irreducible representation are seen along the line of $0.05 \le n \le 0.35$ at $t^{\prime}=0.375$. The type of wave is identified below each figure. It has to be noted that the point at $n = 0.30$ the leading wave is $g^{(8)}$, however we have decided to include the point for completeness reasons. Individual Fermi surfaces are drawn in black.}
\label{pwaveevolution}
\end{figure*}

Our main result, the ground-state phase diagram in the $U\to 0$ limit for the range of density $0 < n < 2$ and the next-nearest-neighbour hopping amplitude  $0\le t^{\prime} \le 0.5$, is presented in Fig.~\ref{phasediagram}. The diagram turns out to be very rich with twelve different states, which we characterize by the number of nodes of the superfluid order parameter, realised with the corresponding symmetries of each of the five irreducible representations. Controlled results at essentially non-zero $U$ for $t^{\prime}=0$ \cite{deng2014emergent} suggest that phases with a high number of nodes, i.e. higher than that in the fundamental mode of the corresponding irreducible representation, tend to disappear as $U$ is increased. In particular, they demonstrate that the $p^{(6)}$ phase, which occupies a significant region in the $U \to 0$ limit (seen in Fig.~\ref{phasediagram} at $t^{\prime}=0, \, n \sim 0.6$) vanishes already at $U=0.08$. Therefore, it is quite possible that this diversity of phases and especially the presence of higher-harmonic states may be a weak-coupling limit artefact. 

We only study positive values of $t^{\prime}$, since for $t^{\prime}<0$ the phase diagram is obtained by reflection symmetry about the point $(t^{\prime}=0, n=1)$, which is due to the mapping of the Hamiltonian onto itself upon replacing all the particles with holes. 
To obtain the phase diagram in Fig.~\ref{phasediagram}, we introduce a grid in the $(t^{\prime}, n)$ plane and find the leading instability by the approach described in Secs.~\ref{subsec:perturbative-treatment}-\ref{subsec:group-theory} at each point of the grid. The horizontal grid step is $\Delta_{t^{\prime}}=0.025$ in regions with multiple phase boundaries in close proximity and $\Delta_{t^{\prime}}=0.05$ everywhere else. The vertical grid step is $\Delta_n=0.05$ everywhere. This high a resolution, which required a substantial computational effort, was necessary to obtain a reliable phase diagram that can be used as a benchmark for further investigations. 

Clearly, the region beneath half filling ($n=1$) is far richer than the region above. This can be related to the presence of the Van Hove line in that part of the diagram. Indeed, most higher-order harmonics are to be found in the vicinity of this line. All irreducible representations have at least one higher-order wave realised over a finite region of the phase diagram. These are however, with the exception of $E_1$, only small regions, mostly at the borders between two or more phases belonging to different irreducible representations. White regions are due to very small values of $\lambda$ ($<10^{-6}$) at which it is difficult to reliably claim which phase is realised. These are, however, not of substantial relevance to high-$T_c$ superconductivity. 

In the following, we discuss the phases classified by their symmetries corresponding to each of the irreducible representations. 

$\mathbf{A_1}$:  The contribution of the fundamental mode $s^{(0)}$ to the vertex $\Gamma^{\text{pp}}$ is negative\cite{brueckner1960, pitaevskii1960, emery1960}, and therefore there is no Cooper instability in the lowest-order $s$-wave channel. Higher harmonics, however, have positive eigenvalues and the corresponding phases are realised at some regions of the parameter space. In particular, it may appear surprising that the $s^{(16)}$ harmonic is dominant in a finite region around $t^{\prime}=0.5$ and $n=0.7$ over the lower-order $s^{(8)}$ harmonic. This may be due to the presence of a second nested line at $t^{\prime}>0.5$.

$\mathbf{A_2}$: The $g^{(8)}$ harmonic dominates at low fillings ($n<0.5$) and intermediate values of hopping ($0.1<t^{\prime}<0.4$). A small region of the $g^{(16)}$ harmonic was found around $t^{\prime}=0.25$ and $n=0.45$. This is a region on the border between the $g^{(8)}$ and $p^{(6)}$ phases and might also be influenced by the $d_{x^2-y^2}^{(4)}$ region at higher doping values. It is thus most likely realised as a frustrated intermediate state on the crossover between those phases. 

$\mathbf{B_1}$: A similar scenario happens at the crossover between $d_{x^2-y^2}^{(4)}$-and $d_{xy}^{(4)}$ regions at $n=1.45$: the crossover between the two is via a strip of the $d_{xy}^{(12)}$ phase. It is interesting to note that the boundary between $d_{x^2-y^2}^{(4)}$ and $d_{xy}^{(4)}$ appears at essentially the same doping independent of the value of $t^{\prime}$. This happens mainly because the shape of the Fermi surface at high values of doping is only weakly dependent on the hoping amplitude. Except for the region of high doping $n>1.45$ another region of $d_{xy}^{(4)}$ exists at small values of $t^{\prime}<0.15$ and fillings $n<0.55$, which is essentially a continuation of the first region reflected at the symmetry point of the phase diagram. Finally a tiny region of $d_{xy}^{(12)}$ was found at $t^{\prime}=0.475$ and $n=0.25$ inside the region dominated by the $E_1$ irreducible representation. It seems to be a consequence of frustration between $p^{(6)}$, $p^{(10)}$ and $p^{(14)}$ as it sits exactly at the boundary between those phases. It is possible that there are multiple such tiny regions spread over the phase diagram that are below our resolution. 

$\mathbf{B_2}$: The $d_{x^2-y^2}^{(4)}$ state dominates in a wide region around half filling. A relatively small region of the $d_{x^2-y^2}^{(12)}$ state was found on the boundary between $d_{x^2-y^2}^{(4)}$ and $s^{(16)}$. As in the case of $s^{(16)}$, we attribute the existence of this $d_{x^2-y^2}^{(12)}$ phase to proximity to the line of Fermi surface nesting at $t^{\prime}>0.5$ discussed above. 

$\mathbf{E_1}$: This irreducible representation displays the richest variety of phases as all first five $p$-type waves are realised in the phase diagram. However, apart from a few exceptions, the geometry of the order parameter is not highly symmetric as in the examples of the $p$-waves in Fig.~\ref{wavesclasses} but with small intervals between some of the nodes, which is generally allowed by the symmetry of $E_1$, often to a degree when it is difficult to identify the exact number of nodes and the crossovers between the corresponding states. The dominant wave in most regions is either $p^{(6)}$ or $p^{(10)}$, and not the fundamental $p^{(2)}$ harmonic, which can be found only in the low density regions. It seems that the $E_1$ phase has the tendency to go towards the $p^{(2)}$ harmonic as $n\rightarrow 0$. Whether this is true for all values of $t^{\prime}$ within the $E_1$ region is unclear as in practice we have only computed the leading instability down to $n =0.05$ which is the minimal value of density within our resolution. 
We see that at low values of $t^{\prime}$ the $p^{(6)}$ state dominates with the exception of a small region of $p^{(10)}$ on the boundary of $g^{(8)}$ and $d_{xy}^{(4)}$. Another region of $p^{(6)}$ phase is at relatively low fillings $0.1<n<0.25$ and $t^{\prime}>0.2$. It transitions smoothly into $p^{(10)}$, turns into $p^{(14)}$ at values of around $n \sim 0.3$. As a typical example, we show the transformation of the $p$-wave order parameter at a fixed $t^{\prime}=0.375$ as the density is increased from $0.05$ to $0.35$ in Fig.~\ref{pwaveevolution}. 
As one approaches the Van Hove line at yet higher values of doping and $t^{\prime}$ the phase diagram becomes patched with both $p^{(6)}$ and $p^{(10)}$ regions present. Interestingly, just below the Van Hove line the $p^{(6)}$ phase becomes clearly dominant again. Above the line there are multiple $p$-wave regions, with $p^{(10)}$ up to about $t^{\prime}>0.38$, followed by a region of  $p^{(14)}$, a small region of $p^{(18)}$, and, finally, again $p^{(6)}$ as the line approaches the FM point at $t^{\prime}\ge 0.47$. Also, at high values of $t^{\prime}>0.4$  and around quarter filling we obtain regions of the high harmonic $p^{(18)}$, similarly to other high-order phases realised in the region ($s^{(16)}$ and $d_{x^2-y^2}^{(12)}$). We suspect that the high order of the leading harmonics is influenced by the aforementioned line of nesting at $t^{\prime} \ge 0.5$. We can also see a clear divide between the influence of the $FM$ and $AFM$ points along the Van Hove line at $t^{\prime}=0.31$: at $t^{\prime}<0.31$ the $d_{x^2-y^2}^{(4)}$ phase dominates, which being a singlet phase corresponds to the symmetry of an AFM-type spin configuration, while at $t^{\prime}>0.31$ the $p^{(6)}$ phase dominates, which being a triplet corresponds to the FM configuration. It is, however, interesting that even close to the FM point $p^{(6)}$ is the leading state instead of $p^{(2)}$.

 \begin{figure*}[!ht]
  \centering
 \raisebox{0.05cm}{\includegraphics[height=5.1cm]{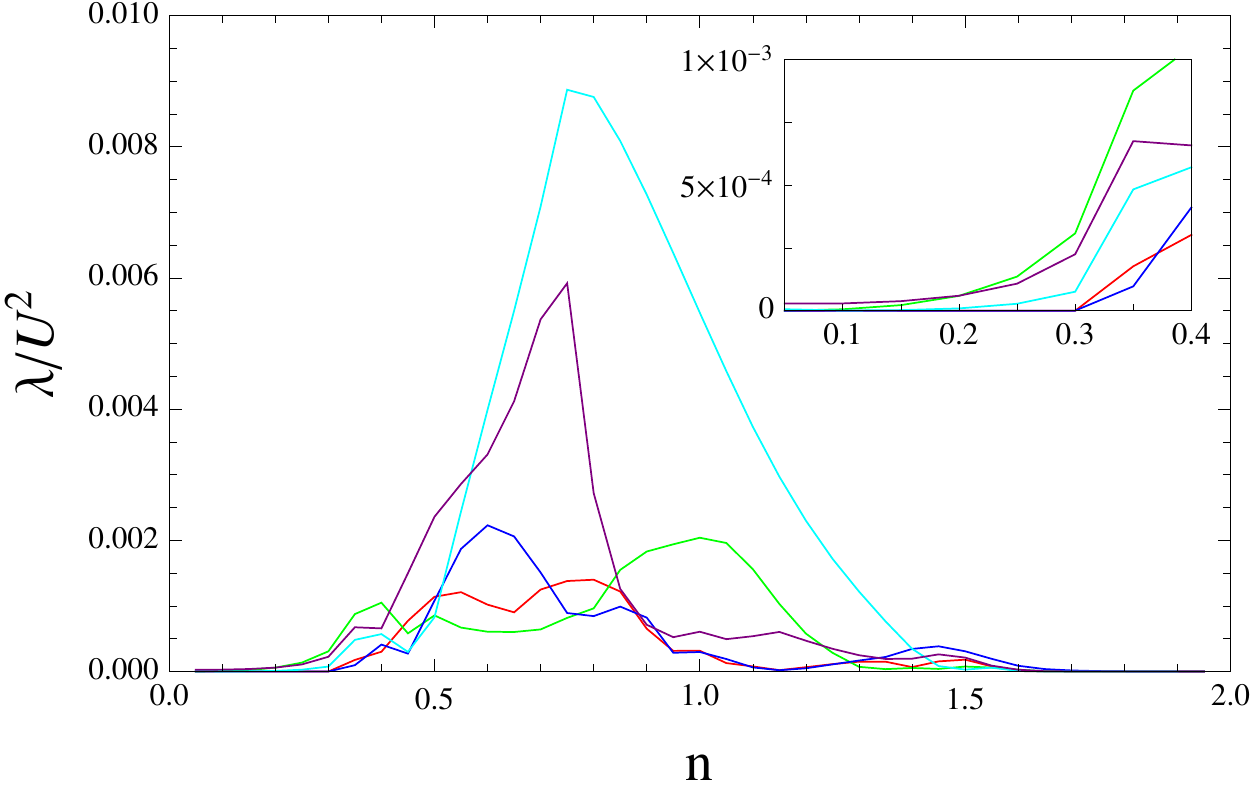}}
\includegraphics[height=5.25cm]{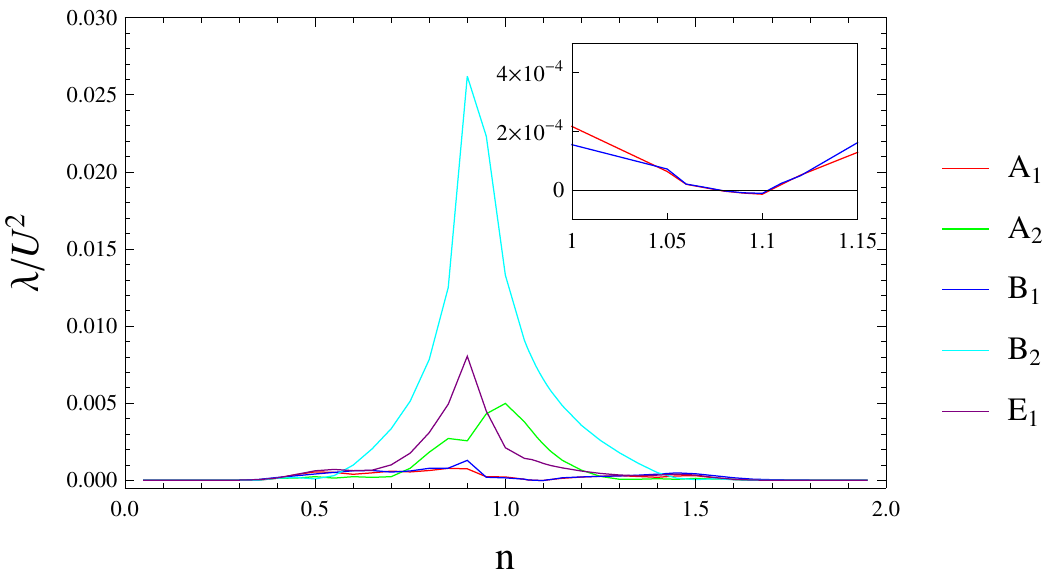}
\vspace{-0.3cm}
\caption{\textit{left}: The leading eigenvalue for each of the irreducible representations as a function of $n$ at $t^{\prime}=0.3$. The highest value of $\lambda$ corresponds to the Van Hove density $n_{VH}\approx 0.726$. The inset is a zoom into the region of low filings $0.05 \le n \le 0.4$. \textit{right}: The leading eigenvalue for each of the irreducible representations as a function of $n$ at $t^{\prime}=0.1$. .The highest value for $\lambda$ corresponds to the Van Hove density $n_{VH}\approx 0.918$. The inset shows negative leading eigenvalues for the $A_1$ and $B_1$ representations in the range $1.08 \le n \le 1.11$}
\label{lambdalines}
\end{figure*}

 \begin{figure}[!ht]
  \centering
\includegraphics[scale=0.75]{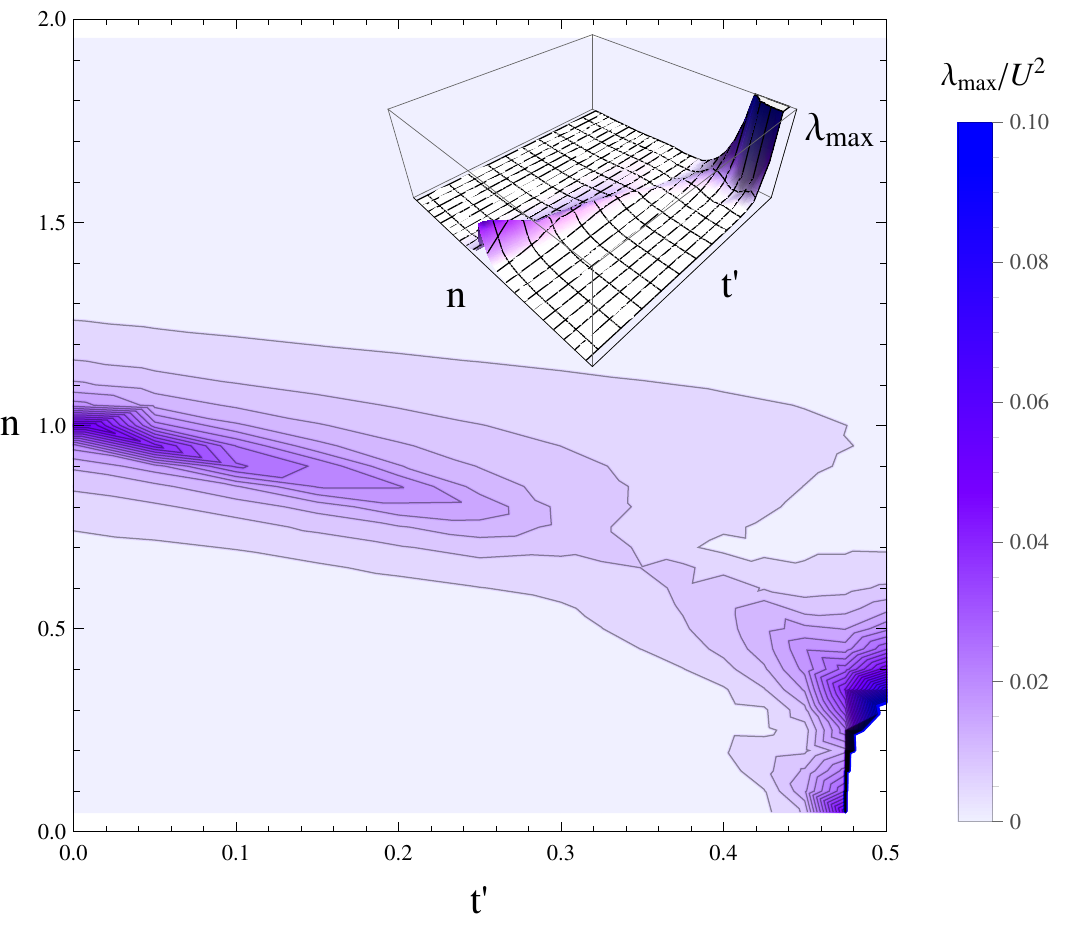}
\vspace{-0.3cm}
\caption{Plot of highest effective coupling constant $\lambda_{max}$ irrespective of type of wave. Inset shows a 3D version of the same plot. Clear maxima can be seen at the FM and AFM point as well as an increase in coupling strength in the vicinity of the Van Hove line.}
\label{effectivecoupling}
\end{figure}


\subsection{Effective coupling strength} \label{subsec:coupling}

The value of the highest effective coupling strength across the parameter space, irrespective of the irreducible representation it belongs to, is plotted in Fig.~\ref{effectivecoupling}. There are two maxima around the FM and AFM points. The maximum in the vicinity of the FM point ($\lambda/U^2 = 13.90$ for $n=0.05$ and $t^{\prime}=0.5$) is much higher than that close to the AFM point  ($\lambda/U^2 = 0.04810$ for $n=1.00$ and $t^{\prime}=0.01$). This is due to the fact that both the particle-hole susceptibility $\chi^{\text{ph}}$ and the density of states diverge at the same vector $q=(0,0)$. However, this does not immediately imply that the critical temperature is also high in this regime --- at low fillings $T_c$ actually becomes low due to the lattice system approaching the continuum limit with a small value of $E_F$. 

Interestingly, at $t^{\prime}=0.5$ the effective coupling constant is relatively large close to quarter-filling $n = 0.5$. This is due to the presence of the line of nesting passing near this point at higher next-nearest-neighbour hoping parameters ($t^{\prime}>0.5$). As an illustration, we computed the effective coupling strength at two points outside of our range of parameters close to this line at $t^{\prime}=0.55$ and $n=\{0.55, 0.60\}$ obtaining $\lambda = \{0.062, 0.0218\}$ respectively, which demonstrates the rapid growth of the effective coupling on approach to the nested point $t^{\prime}=0.5, \, n=0.5$. The effective coupling around $t^{\prime}=0.5, \, n=0.5$ is comparable to the values near the AFM point (to be compared with $\lambda/U^2=0.0481$ at $t^{\prime}=0.01, \, n=1.0$ ). 

A substantial increase in $\lambda$ is clearly seen in the vicinity of the whole Van Hove line. This effect is strongest close to points with nesting, but can be seen even at intermediate values of $t^{\prime}$. In Fig.~\ref{lambdalines} we plot the leading eigenvalues for each of the irreducible representations at two fixed values of the next-nearest-neighbour hopping $t^{\prime}=0.3$ and $t^{\prime}=0.1$ as functions of filling $n$. At $t^{\prime}=0.3$, the density at which the Van Hove singularity crosses the Fermi surface is $n_{VH}\approx 0.726$. Correspondingly, there is a clear peak in the leading $\lambda$ at the Van Hove density, which corresponds to the $d_{x^2-y^2}^{(4)}$ state. At $t^{\prime}=0.1$, the peak of the $d_{x^2-y^2}^{(4)}$ harmonic is also at the density of the Van Hove singularity $n_{VH}\approx 0.918$. One can see from the plot that sub-leading effective couplings can be negative for all harmonics of an irreducible representation. This is true for $A_1$ and $B_1$ representations at densities between $1.08 \le n \le 1.11$.

\begin{figure*}[!ht]
 \begin{center}
\begin{tabular}{ cccccc } 
$A_1$ &$A_2$ & $B_1$ &$B_2$ & $E_1$ &  \\ 
\raisebox{-0.38cm}{\includegraphics[scale=0.28]{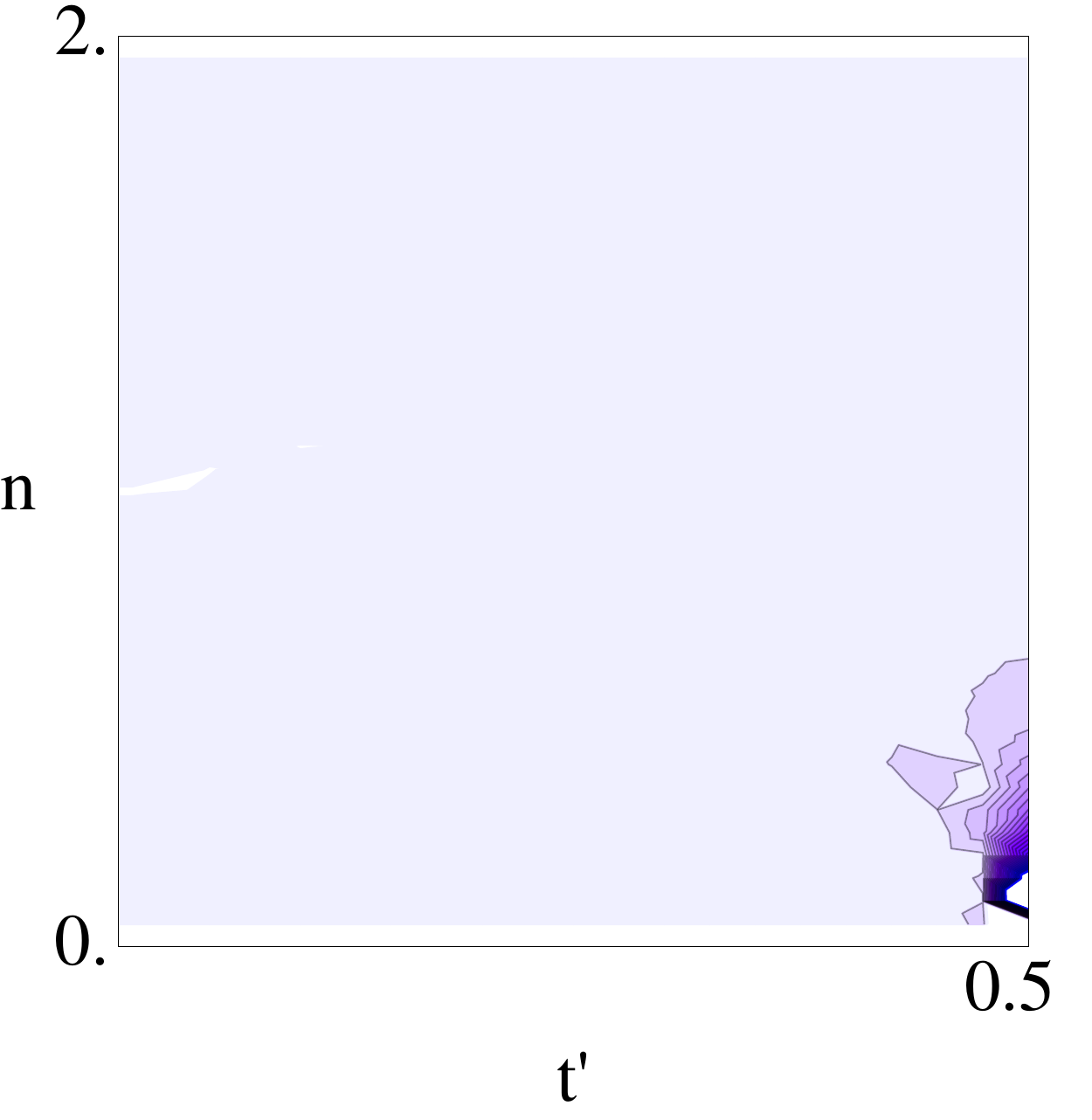}} & \hspace{-0.3cm}
\includegraphics[scale=0.24]{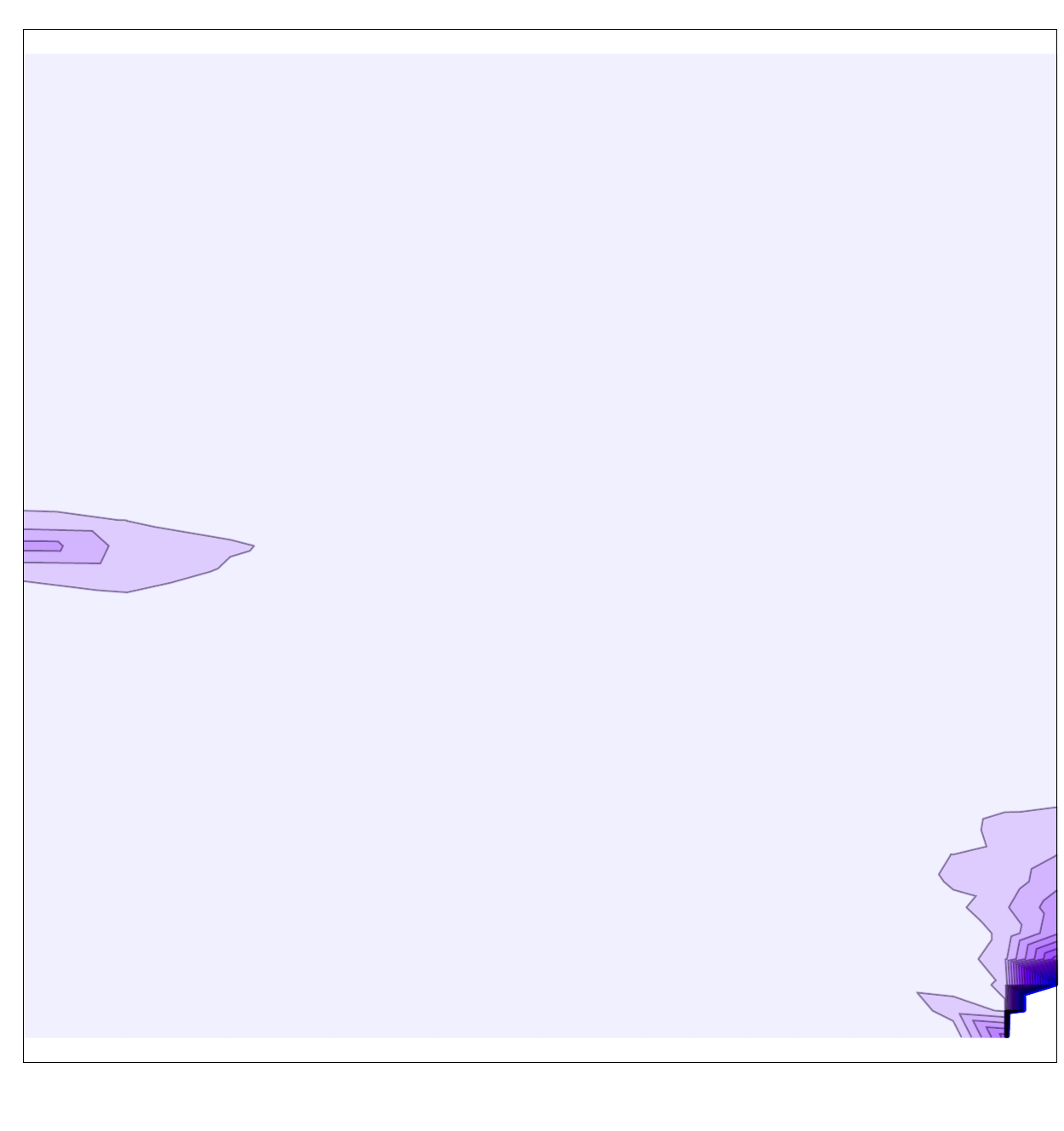} &
\includegraphics[scale=0.24]{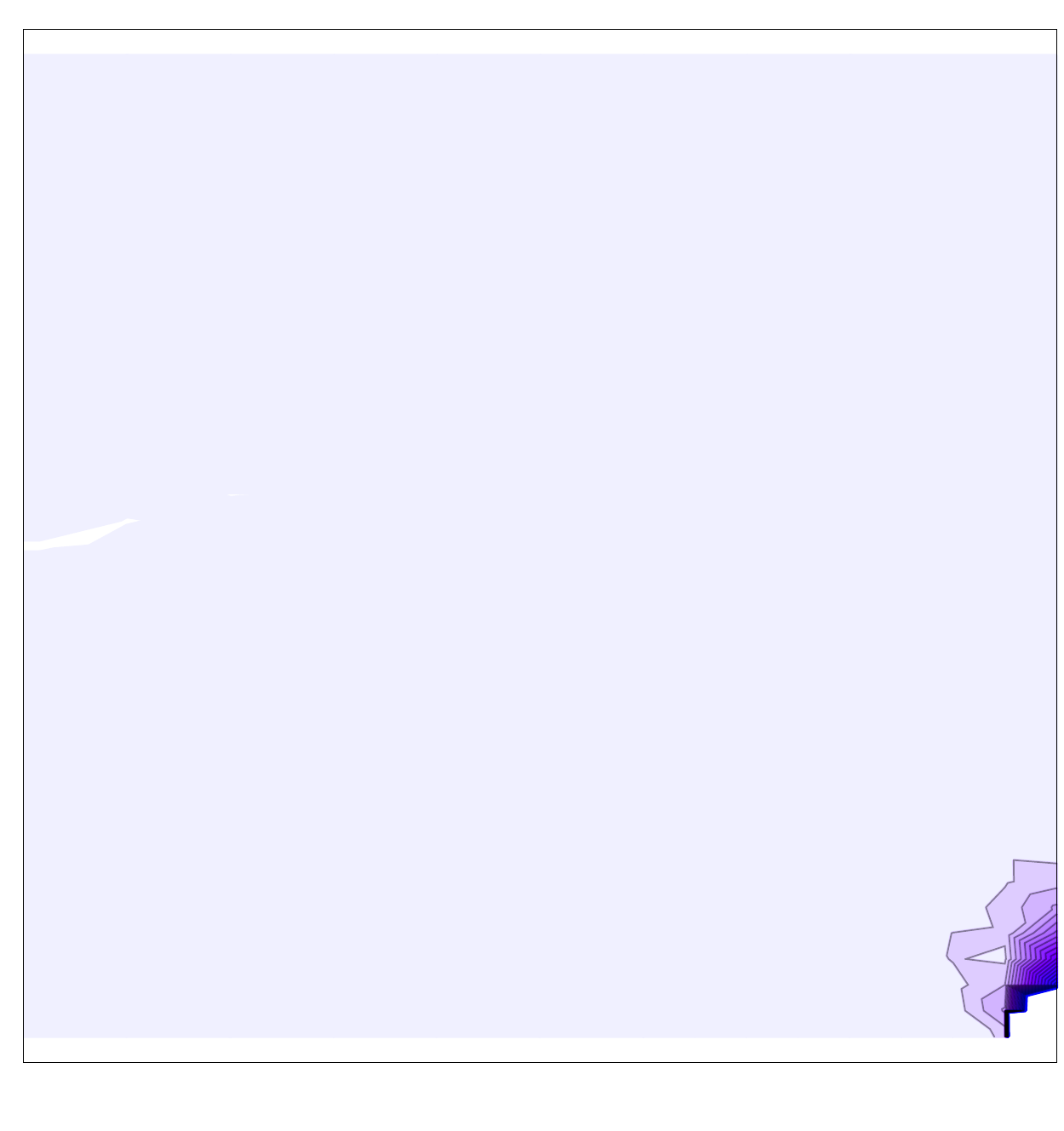} &
\includegraphics[scale=0.24]{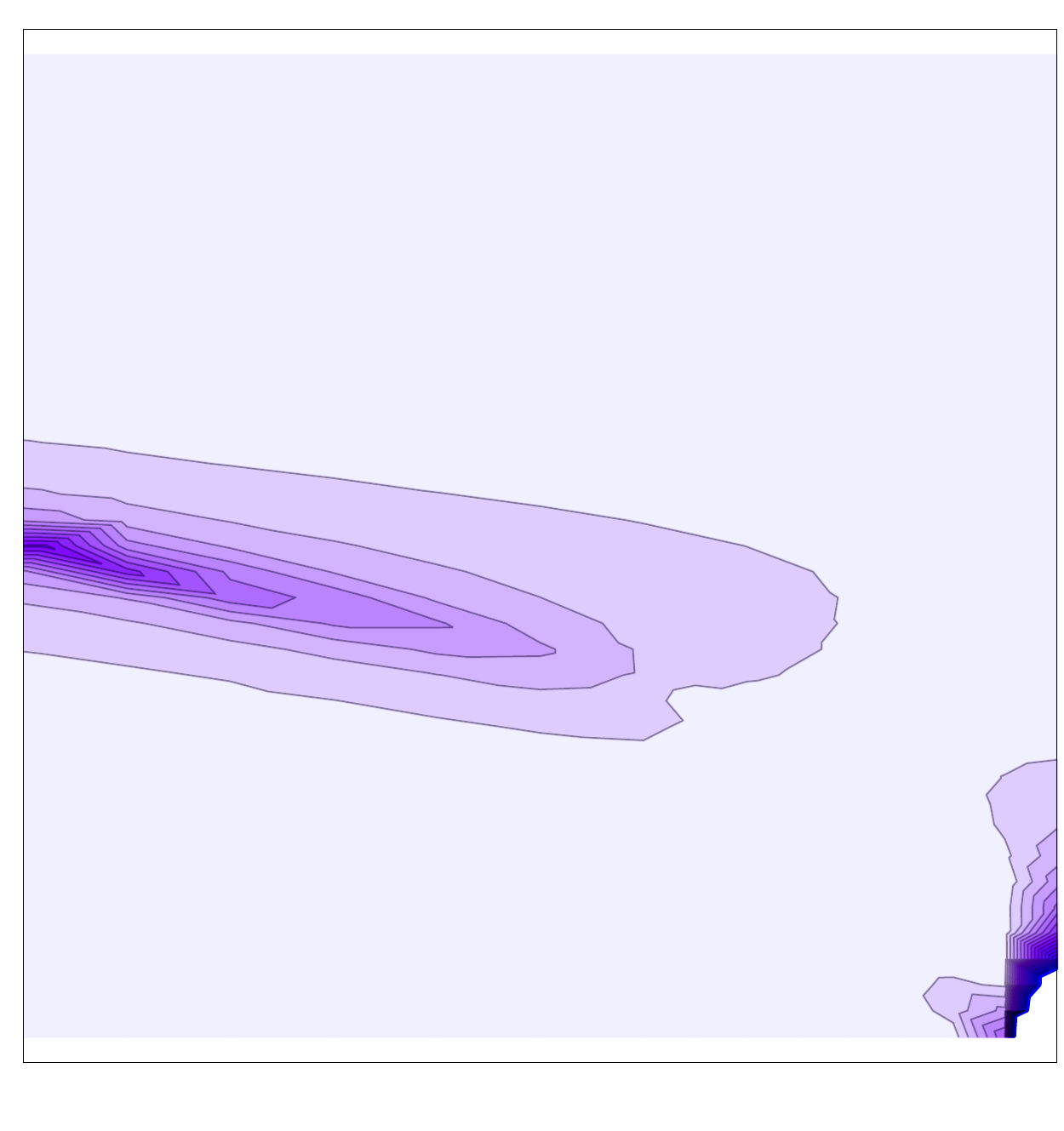} &
\includegraphics[scale=0.24]{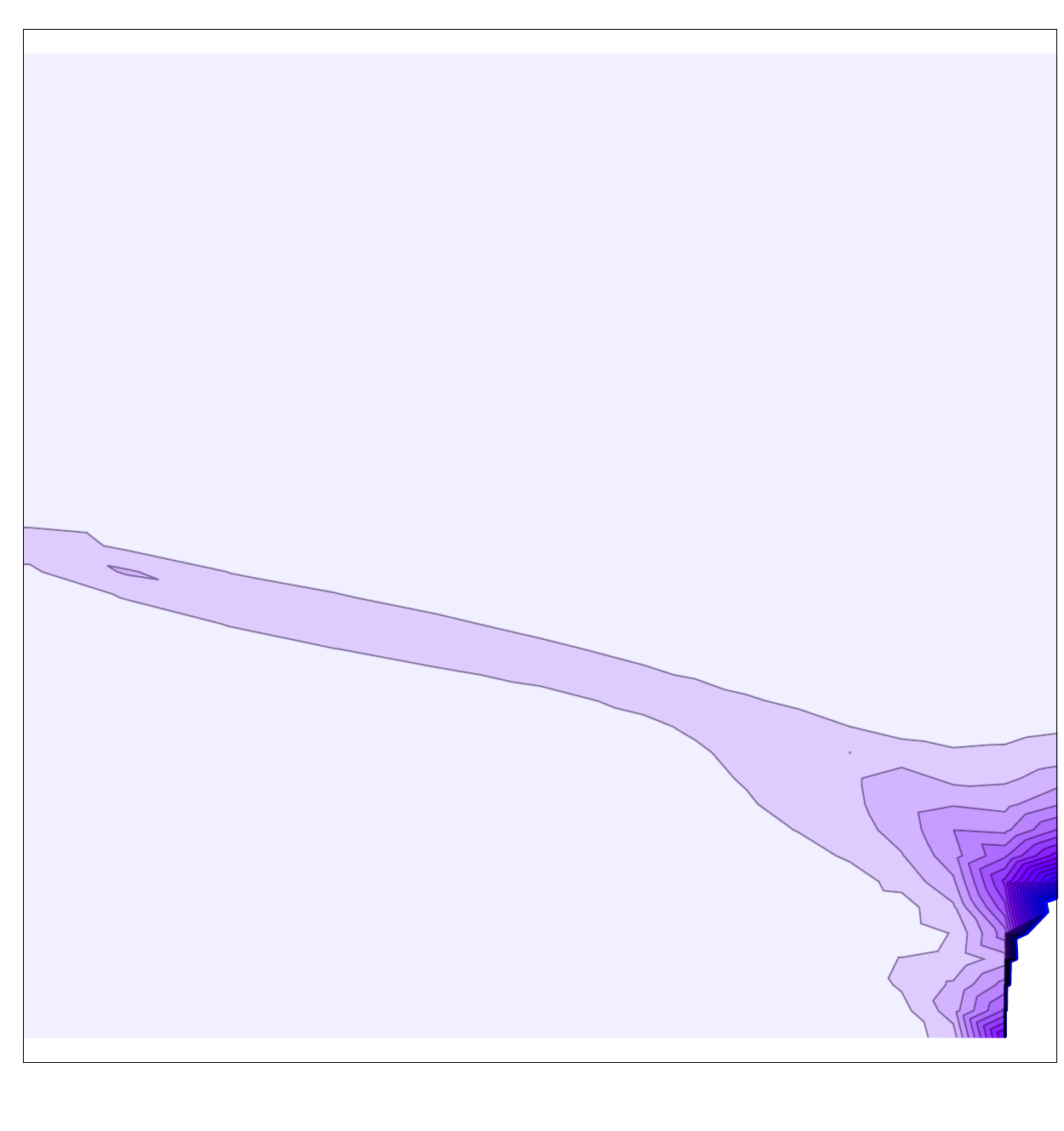} &
\hspace{0.2cm}
\includegraphics[scale=0.26]{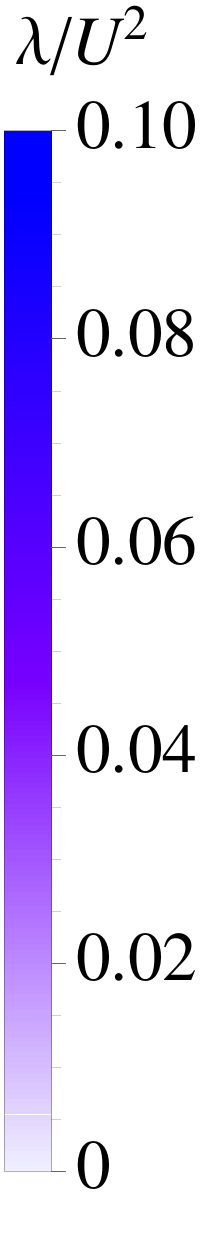} 
\end{tabular}
\end{center}
\vspace{-0.5cm}
\caption{The leading effective coupling constant $\lambda$ for each of the irreducible representations. From left to right: $A_1$, $A_2$, $B_1$, $B_2$, $E_1$. One can see that the AFM point does not affect the effective coupling of $A_1$ and $B_1$ representations, whilst the FM point seems to affect all waves alike. A clear increase in $\lambda$ for $E_1$ can only be seen along the Van Hove line, whilst for $d_{x^2-y^2}^{(4)}$ the coupling strength dies out along the Van Hove line away from the AFM point.}
\label{lambdabyirrep}
\end{figure*}

In Fig.~\ref{lambdabyirrep}, we show contour maps of the leading eigenvalue for each of the irreducible representations in the whole range of parameters. Near the AFM point, only $B_2$ harmonics see a drastic increase in the effective coupling strength, whilst there is also a slight increase for $A_2$. In contrast, the coupling strength for $A_1$ and $B_1$ exhibits a drop near the AFM point. The $E_1$ harmonics have a relatively high eigenvalue along the whole Van Hove line. One can observe a clear increase in coupling strength for all representations near the FM point, which spreads for all the representations except $B_1$ up to quarter filling ($n=0.5$) at $t^{\prime}=0.5$.

\subsection{Comparison to previous work} \label{subsec:previous-work}

Ref. \onlinecite{hlubina1999phase} presents a weak-coupling phase diagram, in a somewhat smaller range of parameters, where the phases are distinguished by their symmetry in terms of the irreducible representations without details of the nodal structure of the order parameter. Our results in Fig.~\ref{phasediagram} (bottom) are mostly consistent with Ref. \onlinecite{hlubina1999phase}. The main difference is the $B_1$ phase ($d_{xy}^{(12)}$ harmonic), which we find between multiple harmonics of the $E_1$-type phase at $t^{\prime}=0.475$ and $n=0.25$, and which is absent in Ref. \onlinecite{hlubina1999phase}. Also, there are two small regions claimed to exist in Ref. \onlinecite{hlubina1999phase} --- a $B_1$-type phase close to the $s^{(16)}$/$E_1$ boundary and an $E_1$-type phase close to the $d_{xy}^{(4)}$ /$g^{(8)}$ --- which we couldn't confirm. Ref.~\onlinecite{hlubina1999phase} does not provide the exact position and extent of the phases, and we do not observe them within our resolution. 

At the level of differentiating between particular harmonics within each irreducible representation the phase diagram becomes far richer. We stress the importance of including high-order harmonics in the analysis because phases of a particular symmetry can be seen to be dominant only at a high expansion order in Eqs.~\ref{harmonics1}-\ref{harmonics2}. In particular, the previously overlooked $B_1$-type phase is realised as $d_{xy}^{(12)}$. Similarly, the $p^{(6)}$ phase at $t^{\prime}=0$ was overlooked in Ref. \onlinecite{raghu2010superconductivity} due to a low number of harmonics allowed in the analysis, but found here as well as in Ref. \onlinecite{hlubina1999phase} and is consistent with recently obtained controlled results at essentially finite $U$~\cite{deng2014emergent}. 

The only other work where an analysis of the effective coupling strength was presented is Ref. \onlinecite{raghu2010superconductivity}. In particular, the paper shows a plot of $\lambda$ as a function of density $n$ at $t^{\prime}=0.3$ (Fig. 4 of Ref. \onlinecite{raghu2010superconductivity}), which, however, is drastically different from our results (let panel of Fig.~\ref{lambdalines}): there is a manifest dip in the effective coupling around the Van Hove density instead of a peak seen in Fig.~\ref{lambdalines}. It appears that the values of $\lambda$ in Ref. \onlinecite{raghu2010superconductivity} are missing a factor of density of states, which essentially changes the result. In fact, we were able to reproduce Fig. 4 of Ref.~\onlinecite{raghu2010superconductivity} by deliberately omitting the density of states from the calculation of $\lambda$. By overestimating the effective coupling by a factor $\sim 40$ close to half filling (and by a factor $\sim 20$ compared to the maximum $\lambda$ that we found at the Van Hove filling), this mistake mislead Ref.~\onlinecite{raghu2010superconductivity} to the incorrect qualitative conclusion that signs of high-$T_c$ superconductivity are already present in the Hubbard model at weak coupling (as also noted in Ref. \onlinecite{kagan2013modern}). Moreover, the $g^{(8)}$ and $p^{(2)}$ states are missing in Fig. 4 of Ref.~\onlinecite{raghu2010superconductivity}, whereas they are the leading instabilities at certain values of filling $n$ within the range shown on the plot.

\section{Conclusions} \label{sec:conclusions}

We have performed a perturbative analysis of the repulsive single-band Hubbard model with next-nearest-neighbour hopping $t^{\prime}$, addressing the Cooper-pairing instability in the Fermi liquid regime in a wide range of $t^{\prime}$ and density $n$. Our results are asymptotically exact in the limit $U \to 0$. We have obtained the ground-state phase diagram and classified phases by their symmetry in terms of the corresponding irreducible representation as well as the nodal structure of the gap function, which resulted in a far richer phase diagram compared to previous works. We have also performed an analysis of the effective coupling strength in the Cooper channel, which controls the superfluid critical temperature. We have observed that the divergence of the density of states at the Fermi surface due to the Van Hove singularity has an influence on both the type of realised superfluid order (the number of nodes in the realised harmonics is anomalously high within the $E_1$ phase around this line) as well as the effective interaction strength (the interaction strength is notably higher in the vicinity of this line). Besides the widely discussed region near the AFM point, we have identified another region with high effective coupling around quarter filling $n=0.5$ and $t^{\prime}=0.5$. This suggests that a detailed study of the model at higher values of next-nearest-neighbour hoppings $t^{\prime}>0.5$ is of substantial interest in the context of high-$T_c$ superconductivity. 

The Fermi-Hubbard model with next-nearest-neighbour hopping could be realised experimentally in optical lattices \cite{dutta2014}, which in principle should allow to obtain the ground-state phase diagram experimentally. Our results provide guidance for future work on detecting high-temperature superconductivity in the model by pointing out the location and extent of regions with high effective coupling at small $U$. The results correct some of the mistakes of previous works, and can serve as a solid foundation for benchmarking advanced computational methods and their application to the model at essentially finite values of coupling $U$. In particular, the phase diagram at essentially finite $U$ can be found by following the evolution of individual phase transition lines from weak coupling by gradually increasing the coupling (as has been done at $t^{\prime}=0$ in Ref.~\onlinecite{deng2014emergent}). Since controlled methods applicable to finite $U$ are extremely computationally expensive, such an approach would drastically improve the efficiency of calculations of the phase diagram by eliminating the need to consider an arbitrary grid of points in the whole parameter space. In this regard, it is worth noting that the landscape of the diagram might change dramatically at finite $U$. In particular, states with a high number of nodes, which are typically found at $U \to 0$ at the boundary between phases belonging to different irreducible representations, could disappear. This is what was found to happen in Ref.~\onlinecite{deng2014emergent} at $t^{\prime}=0$, where the slab of the $p^{(6)}$ phase at the boundary between $d_{xy}^{(4)}$ and $d_{x^2-y^2}^{(4)}$ vanished already at $U=0.08$.


\section{Acknowledgements}
The authors would like to thank N. Prokofiev, B. Svistunov and S. Nikolaev for valuable discussions. This work was supported by the Simons Collaboration on the Many Electron Problem, National Science Foundation under the grant PHY-1314735. Y. Deng and X.W. Liu thank for the support of the National Natural Science Foundation of China under Grant No.11275185, and the Fundamental Research Funds for the Central Universities under Grant No.2340000034.

\bibliographystyle{apsrev4-1}
\bibliography{squarehubbard}

\end{document}